%% file: bare_jrnl_compsoc.tex
\documentclass[10pt,journal,compsoc]{IEEEtran}
\usepackage{caption}
\usepackage{url}
\usepackage{lipsum}
\usepackage{algorithmic}
\usepackage[linesnumbered,lined,boxed,commentsnumbered]{algorithm2e}
\usepackage{multicol}
\usepackage[utf8]{inputenc}
\usepackage{amssymb}
\usepackage{subfigure}
\usepackage{color}
\usepackage{url}
\usepackage{multirow}
\usepackage{epsfig}
\usepackage{epstopdf}
\usepackage{graphicx}
\usepackage{subfig}
\usepackage{amssymb}
\usepackage[figuresright]{rotating}
\usepackage{float}

\usepackage{ragged2e}

\ifCLASSOPTIONcompsoc
  \usepackage[nocompress]{cite}
\else
  \usepackage{cite}
\fi
\ifCLASSINFOpdf
\else
\fi
\begin{document}
\title{Privacy-preserving targeted mobile advertising: A Blockchain-based framework for mobile ads}

\author{Imdad Ullah,~\IEEEmembership{Member, IEEE,} Salil S. Kanhere,~\IEEEmembership{Senior Member, IEEE,}
        and Roksana Boreli% <-this % stops a space
\IEEEcompsocitemizethanks{\IEEEcompsocthanksitem I. Ullah is with the College of Computer Engineering and Sciences, Prince Sattam bin Abdulaziz University, Al-Kharj 11942, Saudi Arabia.\protect\\
E-mail: i.ullah@psau.edu.sa
\IEEEcompsocthanksitem S. S. Kanhere is with UNSW Sydney, Australia.\protect\\
E-mail: salil.kanhere@unsw.edu.au
\IEEEcompsocthanksitem R. Boreli was with CSIRO Data61 Sydney, Australia.\protect\\
E-mail: roksana@tmppbiz.com}% <-this % stops an unwanted space
}

\IEEEtitleabstractindextext{%
\justify
\begin{abstract}
The targeted advertising is based on preference profiles inferred via relationships among individuals, their monitored responses to previous advertising and temporal activity over the Internet, which has raised critical privacy concerns. In this paper, we present a novel proposal for a Blockchain-based advertising platform that provides: a system for privacy preserving user profiling, privately requesting ads from the advertising system, the billing mechanisms for presented and clicked ads, the advertising system that uploads ads to the cloud according to profiling interests, various types of transactions to enable advertising operations in Blockchain-based network, and the method that allows a cloud system to privately compute the access policies for various resources (such as ads, mobile user profiles). Our main goal is to design a decentralized framework for targeted ads, which enables private delivery of ads to users whose behavioral profiles accurately match the presented ads, defined by the ad system. We implement a POC of our proposed framework i.e. a \textit{Bespoke Miner} and experimentally evaluate various components of Blockchain-based \textit{in-app} advertising system, implementing various critical components; such as, evaluating user profiles, implementing access policies, encryption and decryption of users' profiles. We observe that the processing delay for traversing policies of various tree sizes, the encryption/decryption time of user profiling with various key-sizes and user profiles of various interests evaluates to an acceptable amount of processing time as that of the currently implemented ad systems.
\end{abstract}

}

\maketitle

\IEEEdisplaynontitleabstractindextext

% \fi
\IEEEpeerreviewmaketitle

\section{Introduction} \label{section-introduction}
\input{introduction}

%\section{System Model} \label{section-obfuscation}
\section{Problem Formulation} \label{problem-formulation}
\input{background}

\section{System Components} \label{system-model}
\input{system-model}

\section{Ad-Block Structure} \label{ad-block-structure}
\input{ad-block-structure}

\section{Ad-Block Operations} \label{ad-block-operations}
\input{adblock-operation}

\section{Performance Evaluation} \label{evaluation}
\input{evaluation}

\section{Related Work} \label{related-work}
\input{related-work}

\section{Conclusion} \label{conclusion}
\input{conclusion}

%\bibliographystyle{ieeetr}
%\bibliography{references}

\begin{IEEEbiography}[{\includegraphics[width=1in,height=1.25in,clip,keepaspectratio]{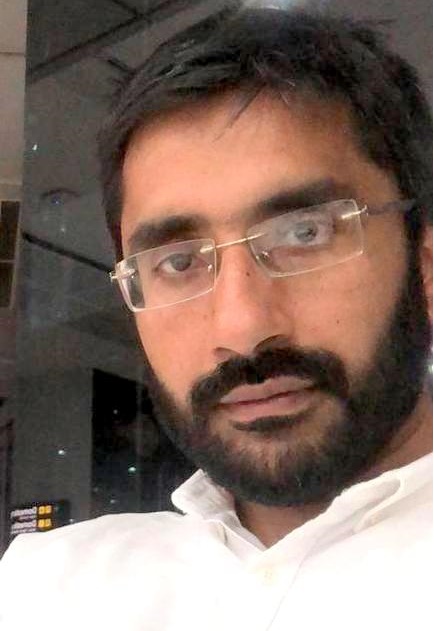}}]{Imdad Ullah} has received his Ph.D. in Computer Science and Engineering from The University of New South Wales (UNSW) Sydney, Australia. He is currently an assistant professor with the College of Computer Engineering and Sciences, PSAU, Saudi Arabia. He has served in various positions of Researcher at UNSW, Research scholar at National ICT Australia (NICTA)/Data61 CSIRO Australia, NUST Islamabad Pakistan and SEEMOO TU Darmstadt Germany, and Research Collaborator at SLAC National Accelerator Laboratory Stanford University USA. He has research and development experience in privacy preserving systems including private advertising and crypto-based billing systems. His primary research interest include privacy enhancing technologies; he also has interest in Internet of Things, Blockchain, network modeling and design, network measurements, and trusted networking.
\end{IEEEbiography}

\begin{IEEEbiography}[{\includegraphics[width=1in,height=1.25in,clip,keepaspectratio]{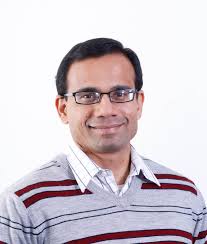}}]{Salil S. Kanhere} (Senior Member, IEEE) received the M.S. and Ph.D. degrees from Drexel University, Philadelphia. He is currently a Professor of Computer Science and Engineering with UNSW Sydney, Australia. His research interests include the Internet of Things, cyberphysical systems, blockchain, pervasive computing, cybersecurity, and applied machine learning. He is a Senior Member of the ACM, an Humboldt Research Fellow, and an ACM Distinguished Speaker. He serves as the Editor in Chief of the Ad Hoc Networks journal and as an Associate Editor of the IEEE Transactions On Network and Service Management, Computer Communications, and Pervasive andMobile Computing. He has served on the organising committee of several IEEE/ACM international conferences. He has co-authored a book titled \textit{Blockchain for Cyberphysical Systems}.
\end{IEEEbiography}

\begin{IEEEbiography}[{\includegraphics[width=1in,height=1.25in,clip,keepaspectratio]{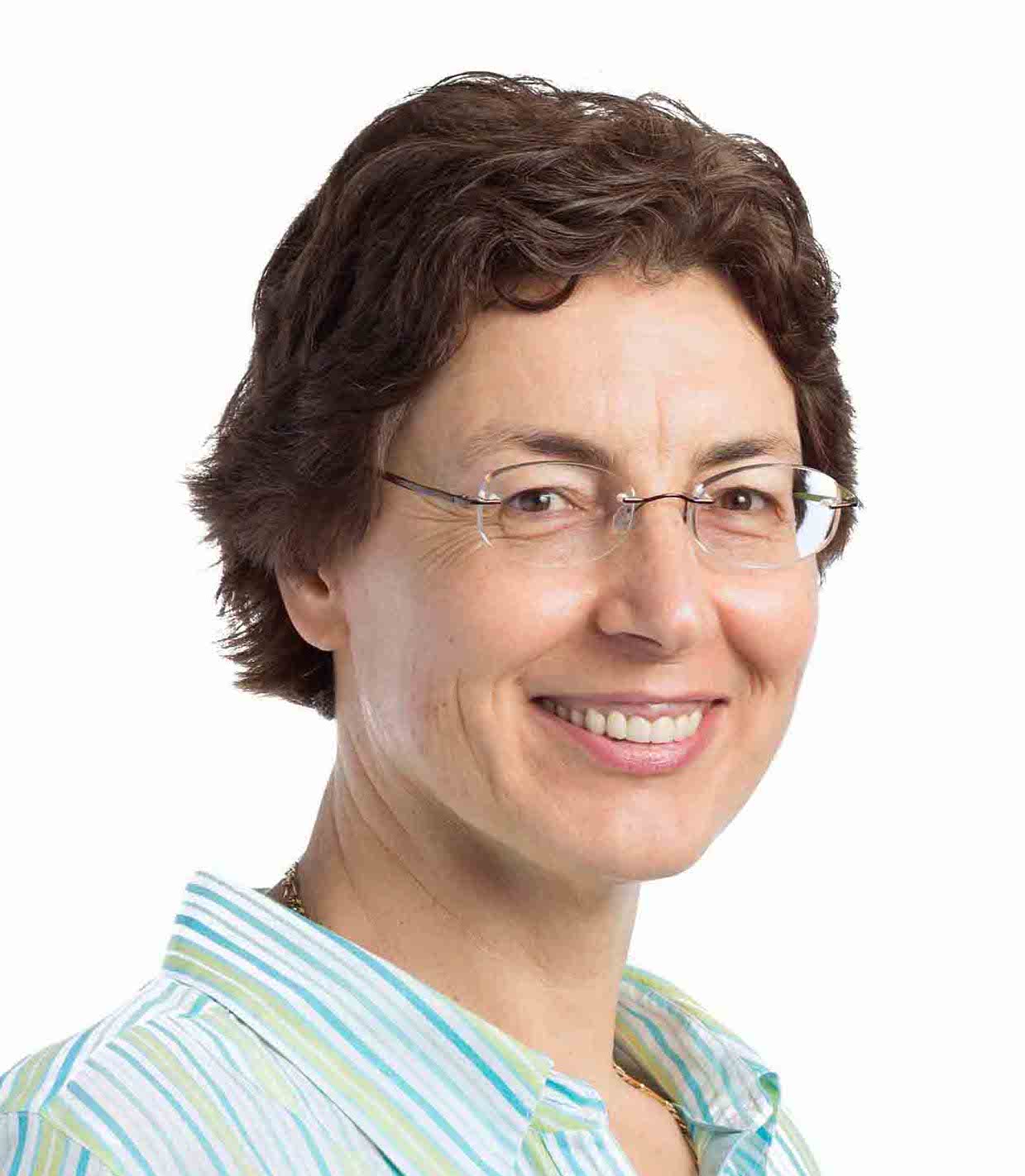}}]{Roksana Boreli} has received her Ph.D in Communications from University of Technology, Sydney, Australia.  She has over 20 years of experience in communications and networking research and in engineering development, in large telecommunications companies (Telstra Australia, Xantic, NL) and research organisations. Roksana has served in various positions of Engineering manager, Technology strategist, Research leader of the Privacy area of Networks research group in National ICT Australia (NICTA)/CSIRO Data61 and CTO in a NICTA spinoff 7-ip. Her primary research focus is on the privacy enhancing technologies; she also maintains an interest in mobile and wireless communications.
\end{IEEEbiography}


%\vfill


\end{document}

%% file: introduction.tex
Mobile advertising ecosystem is one of the most successful markets, in recent years, with billions of mobile devices, including smartphones, tablets, and computer tablets, and millions of mobile applications (apps) registered in various app platforms, such as Google Play Store, Apps Store etc. The advertising companies enable user tracking within apps, and hence profiling users, where user's personal information plays an important role in ads targeting. However, significant privacy concerns are associated with it, as suggested in a number of studies \cite{mamais2019privacy, ullah2017enabling, liu2016privacy, wang2020aiming, ullah2020protecting, tchen2014}. Other studies \cite{tsang2004consumer, merisavo2007empirical} indicate, unless consumers have specifically consented to it, that they have a direct relationship between consumer attitude and their behavior, i.e. consumers' trust in privacy of mobile advertising is positively related to their willingness to accept mobile advertising. The American ad industry, implemented self-regulated by implementing AdChoices\footnote{Founded in 2010. \url{https://optout.aboutads.info/?c=2&lang=EN}} program, which states that consumer could \textit{opt-out} of \textit{targeted} advertising via online choices to control ads from other networks. However, another study \cite{johnson2020consumer} examines that \textit{opt-out} users cause 52\% less revenue (hence presents less relevant ads and lower click through rates) than those users who allow \textit{targeted} advertising. In addition, the authors noted that these ad impressions were only requested by 0.23\% of American consumers.

Hence, the purpose of \textit{targeted} advertising is to deliver most relevant ads to achieve better view/click through rates and without exposing users' private information to the advertising companies and third party publishers/ad networks. Prior research works show the extent to which consumers are effectively tracked by a number of third parties and across multiple apps \cite{razaghpanah2018apps}, mobile devices leaking Personally Identifiable Information (PII) \cite{elsabagh2020firmscope, ren2016recon}, apps accessing  user's  private/sensitive information through well defined APIs \cite{verderame2020reliability}, inference attacks based on monitoring ads \cite{tchen2014} and other data platform such as eXelate\footnote{\url{https://microsites.nielsen.com/daas-partners/partner/exelate/}}, BlueKai\footnote{\url{https://www.oracle.com/corporate/acquisitions/bluekai/}}, and AddThis\footnote{\url{https://www.addthis.com/}} that collect, enrich and resell cookies.

There are other works on protecting user's privacy by obfuscating user profiles\footnote{A large scale investigation of obfuscation use where authors analyze 1.7 million free Android apps from Google Play to detect various obfuscation techniques is presented in \cite{wermke2018large}.} \cite{ullah2014profileguard, ullah2020protecting} and based on security techniques \cite{ullah2017enabling}, enabling user privacy \cite{balebako2014improving, balebako2014privacy} by suggesting design changes to Android location API based on laboratory study \cite{jain2014should} to understand developer's behavior. Other privacy preserving advertising systems use a mix of cryptography techniques and obfuscation mechanisms, e.g., \cite{guha2011privad, toubiana2010adnostic} and \textit{targeted} mobile coupon delivery scheme using Blockchain \cite{gu2018secure}. We note that the majority of the private advertising systems have been proposed for browser based ads \cite{kenthapadi2019random, beigi2019protecting, starov2018privacymeter, laoutaris2018method} whereas only few works address the \textit{in-app} \textit{targeted} ads \cite{gu2018secure, rafieian2020targeting, ullah2014profileguard, ullah2017enabling}. Our focus in this work is to protect individual's privacy by decoupling all direct communications between apps, analytics companies, which they use for \textit{targeted} advertisements, advertising systems (including third party) for their activities (both for web and apps activity) using Blockchain technology.

Blockchain is based on a distributed ledger of transactions, shared across the participating entities, and provides auditable transitions \cite{kosba2016hawk}, where the transactions are verified by participating entities within operating network. Among the participating entities; the \textit{Miner} is a special node responsible for generating transactions, adding them to the pool of pending transactions and organizing into a \textit{block} once the size of transactions reaches a specific \textit{block size}. The process of adding a new block to the Blockchain is referred to as \textit{mining} and follows a consensus algorithm, such as Proof of Work (POW) \cite{vukolic2015quest} and Proof of Stake (POS) \cite{wood2014ethereum}, which ensures the security of Blockchain against malicious (\textit{Miner}) users. The participating entities use the \textit{Public-Private Key} pair that is used to achieve the anonymity \cite{dorri2017blockchain}. Among various salient features of Blockchain, i.e. irreversible, auditable, updated near real-time, chronological and timestamp, which, in addition, disregards the need of a central controlling authority; thus making it a perfect choice for restricting communication between the mobile apps and the analytics/ad companies and keeping individual's privacy.

%Blockchain is based on a distributed ledger of transactions, shared across the participating entities, and provides auditable transitions \cite{kosba2016hawk}, where the transactions are verified by participating entities within operating network. Among the participating entities; the \textit{Miner} is a special node responsible for generating transactions, adding them to the pool of pending transactions and organizing into a \textit{block} once the size of transactions reaches a specific \textit{block size}. The process of adding a new block to the Blockchain is referred to as \textit{mining} and follows a consensus algorithm, such as Proof of Work (POW) \cite{vukolic2015quest} and Proof of Stake (POS) \cite{wood2014ethereum}, which ensures the security of Blockchain against malicious (\textit{Miner}) users. The participating entities use the Public-Private Key pair that is used to achieve the anonymity \cite{dorri2017blockchain}. Among various salient features of Blockchain, i.e. irreversible, auditable, updated near real-time, chronological and timestamp, which, in addition, disregards the need of a central controlling authority; thus making it a perfect choice for restricting communication between the mobile apps and the analytics/ad companies and keeping individual's privacy.

Our main contributions in this paper include the following. We investigate the relation between the mobile apps usage and the resulting user profiles derived by a major analytics network i.e., Google AdMob. We mainly describe three states of the user profiles i.e. the user profiles that are established during when no interests are derived by the analytics services i.e. \textit{profile establishment} process, the state of the profiles once apps are used that would originate interests other than the existing set of apps i.e. \textit{profile evolution} process, and \textit{stable states} of the profiles where subsequent use of existing set of apps have no effect over the profiles. We also determine the broad profiling rules during various processes of utilising android apps: that profiles represent an aggregation of interests generated by use of individual apps, that there is a fixed mapping of apps to sets of interests and that a minimum level of activity (in a period of up to 72h) is required to build a \textit{stable} user profile. We correspondingly develop an analytical model for depicting various profiling states and rules to represent profiling processes.

Following, we design a framework for secure user profiling and Blockchain-based \textit{targeted} advertising system for \textit{in-app} mobile ads and the design of various critical components of advertising system, as follows: the \textbf{1}. Ads Placement Server (\texttt{APS}) for evaluating ads according to various profile interests, hashing ads and sending hashed profiles along with corresponding ads to \texttt{CS} \textbf{2}. Cloud System (\texttt{CS}) for storing hashed profiles along with corresponding ads sent by \texttt{APS}, in addition, maintaining a copy of user profiles \textbf{3}. Cluster Head (\texttt{CH}) for processing \textit{ad-block} requests, \textit{access control policy} for various operations, such as storing user profile etc., and fetching ads from \texttt{CS} according to user profiles \textbf{4}. System App (\textit{Miner}) that sits on user's mobile device and calculates user's profiles, sends hashed user profile to \texttt{CH}, uploads user profiles on \texttt{CS}, and requests and delivers ads to mobile app \textbf{5}. Billing Server (\texttt{BS}) for calculating \textit{billing} for presented and clicked ads with the use of Cryptocurrency for ensuring privacy, secure payment and for compatibility with the underlying proposed advertising system over Blockchain \textbf{6}. \textit{Access Control Policy} implemented in \textit{Miner} for holding an \textit{access control list} that allows apps regarding evaluation of different operations e.g. to check whether an app is allowed to request for ads, participate in \textit{billing} etc., and on \texttt{CS} to control access of \textit{ad-request} coming from various \textit{Miners}. The \textit{access policy} is implemented as a \textit{policy tree} that consists of matching functions for allowing access to various operations or resources and their corresponding actions.

\sloppy We design the structure for various transactions for an \textit{ad-request/response block} for operating within the Blockchain network e.g. block structure of an \texttt{Ad-Block-Header-Request}, \texttt{Ad-Block-Header-Response}, and \texttt{Ad-Block-Structure}. We design various algorithms that specify comprehensive operations of various components of the advertising system. Furthermore, we discuss various design requirements in terms of privacy, reduced overhead over (user) client, compatibility of \textit{billing} with Cryptocurrency, and achieving system efficiency and component's compatibility within the Blockchain network.

We develop an Android-based Proof of Concept (POC) implementation i.e. a \textit{Bespoke Miner} for evaluating various components of Blockchain-based \textit{in-app} advertising system, such as, generating \textit{public-private key} pair, evaluating user profiles,  encrypting user profile or decrypting the ads received from \texttt{CS}, implementing \textit{access policy}, and its interaction with the \texttt{CS}. We design a detailed experimental setup that consists of \textit{Bespoke Miner} and \texttt{CS} (a local machine for correspondence with the \textit{Bespoke Miner}), following, we utilise real parameter settings obtained by extensive our real-time experiments \cite{ullah2014characterising, ullah2014profileguard} (and further verified that these parameters still hold) that include various sizes of an ad, set of interests with a user profile etc. We calculate the processing delay for generating \textit{public-private key} pair, encryption/decryption time of user profiling with various key-sizes and user profiles, and processing delay for traversing \textit{policies} of various sizes. We observe that, using various $512, 1024, 2048, 4096$ bits sizes of \textit{public-private key} pairs, the $Min$ processing time ranges from 0.005sec to 0.30sec.

Following, we calculate encryption and decryption time for user profiles with interests, randomly chosen up to 20 interests with an $Avg$ and $St.Dev$ of respectively 10.53 and 5.88 interests, using $1024, 2048, 4096, 8192$ bits sizes of \textit{public-private key} pairs and evaluate the processing delays of $Avg, Min, Max, St.Dev$. We observe that, for instance, the $Min$ and $Max$ profile encryption times with $(1024, 8192)$ bits key sizes are respectively (0.1msec, 1.2msec) and (2.2msec, 16.3msec), with an $Avg$ and $St.Dev$ of (0.4msec) each and (6.0msec, 3.2msec). In addition, the decryption time evaluates to (0.001sec, 0.005mse) and (0.35sec, 1.67sec) with $Avg$ and $St.Dev$ of (0.002sec, 0.0009sec) and (0.8sec, and 0.36sec). Based on our observations \cite{ullah2020protecting, ullah2017enabling}, we note that this is an (equivalent) acceptable amount of processing time to that of when a mobile app requests ad in the beginning of an app launch or during the app usage session.

%Following, we calculate the encryption and decryption time for user profiles with interests, randomly chosen up to 20 interests with an $Avg$ and $St.Dev$ of respectively 10.53 and 5.88 interests, using $512, 1024, 2048, 4096$ bits sizes of \textit{public-private key} pairs and evaluate its performance in terms of processing delays of $Avg, Min, Max, St.Dev$. We observe that, for instance, the $Min$ and $Max$ profile encryption time with $(1024, 8192)$ bits keys size are respectively (0.1msec, 1.2msec) and (2.2msec, 16.3msec), with an $Avg$ and $St.Dev$ of (0.4msec) each and (6.0msec, 3.2msec) respectively with $1024$ and $8192$ bits keys size. In addition, the decryption time respectively with $(1024, 8192)$ bits keys size, the $Min$ and $Max$ evaluates to (0.001sec, 0.005mse) and (0.35sec, 1.67sec) with $Avg$ and $St.Dev$ of (0.002sec, 0.0009sec) and (0.8sec, and 0.36sec). Based on our observations \cite{ullah2014characterising, ullah2017enabling}, we note that this is an (equivalent) acceptable amount processing time to that of when a mobile app requests ad in the beginning of an app launch or during the app usage session.

\sloppy Rest of the paper is organised as follows: Section \ref{problem-formulation} presents detailed discussion over the profiling process and formulates the problem. Section \ref{system-model} presents various components of the proposed framework for secure user profiling and Blockchain-based \textit{targeted} advertising system for \textit{in-app} mobile ads. In Section \ref{ad-block-structure}, we present the \texttt{Ad-Block-Structure} for an \textit{ad-request/response} for operating within the Blockchain network. Details of \texttt{Ad-Block} operations is presented in \ref{ad-block-operations}, followed by experimental setup, including details of POC implementation for bespoke \textit{Miner}, and experimental evaluation in Section \ref{evaluation}. Related work is summarised in Section  \ref{related-work} and conclusion of our work is given in Section \ref{conclusion}.

%% file: background.tex
We formalise the system model that consists of the applications' profiles, interests' profiles and the conversion of resulting profiles by use of applications in an app profile. In particular, we provide the insights of \textit{establishment} of \textit{Interests profiles} by individual apps in the \textit{Apps profiles} and then show how the profiles \textit{evolve} when some apps other than the initial set of apps are utilised. %We then detail the evaluation of ads targeting by first describing the experimental setup, the procedure for collecting ads and their related traffic, and then detail the criteria for classification of ads into different categories.

\subsection{Profiling}\label{user-profiling}
An app marketplace includes a set of $\mathcal{A}$ mobile apps, that are commonly (e.g., in Google Play or in the Apple App store) organised within different categories. We denote an app by ${a_{i,j}}$, $i=1,...,A_j$, where $A_j$ is the number of apps that belong to an app category $\Phi _j$, $j=1,..., \tau $ and $\tau $ is the number of different categories in the marketplace.

A user may be characterised by a combination of apps installed on their mobile device(s), comprising a subset $S_a$ of the apps set $\mathcal{A}$. An \textit{Apps profile} ${K_a}$ can therefore be defined as:

\begin{equation}
%{K_a} = \left\{ {\left\{ {{a_{i,j}},{\Phi _j}} \right\}:{a_{i,j}} \in {S_a}} \right\}
{K_a} = \left\{ {\left\{ {{a_{i,j}},{\Phi _j}} \right\}:{a_{i,j}} \in {S_a}} \right\}
\end{equation}

Analytics companies (e.g., Google or Flurry) profile users by defining a set of profile interests $\mathcal{G}$, i.e., characteristics that may be assigned to users. Interests are commonly grouped into categories, with interests ${g_{k,l}}$, $k=1,...,G_l$, where $G_l$ is the number of interests that belong to an interest category $\Psi_l$,  $l=1,..., \epsilon $.  $\epsilon $ is the number of interest categories defined by the analytics company. Table \ref{table:notations} shows notations used in this work.

Profiling characterises the user by the combination of their interests, i.e., by a subset $S_g$ of the full interest set $\mathcal{G}$. Google profile interests\footnote{Google profile interests are listed in \url{https://adssettings.google.com/authenticated?hl=en}, displayed under the 'How your ads are personalized', other options and Google services can also be verified on Google Dashboard \url{https://myaccount.google.com/dashboard?hl=en}.} are grouped, hierarchically, under vaiours interests categories, with specific interests. An \textit{Interests profile} ${I_g}$ for a user can therefore be defined as:

\begin{equation}\label{eq:initial-interes-profile}
{I_g} = \left\{ {\left\{ {{g_{k,l}},{\Psi _l}} \right\}:{g_{k,l}} \in {S_g}} \right\}
\end{equation}

We note that although various types of information may be used to generate user profiles, here we focus on the interests derived from the usage of installed apps. Note that we will use the \textit{Apps profile} ${K_a}$ and \textit{Interests profile} ${I_g}$ respectively for calculating the \textit{Contextual} and \textit{Targeted} ads.

 %s i.e. \ref{subsection:profileEstablishment}, \ref{subsection:profileEvolution}, and \ref{subsection:overlProcess}

\begin{table*}[t]
\begin{center} %\scalebox{0.7}{
{\begin{tabular}{c|l}
\hline
 Symbols& Description\tabularnewline
\hline
${\mathcal{A}}$& Set of apps in a marketplace \tabularnewline
\hline
$\tau$ & Number of app categories, $\Phi_j$ is a selected category, $j=1,...,\tau$ \tabularnewline
\hline
 ${S_a}$& Subset of apps installed on a user's mobile device\tabularnewline
\hline
 ${a_{i, j}}$& An app ${{a_{i,j}} \in A}$,  $i=1,...,A_j$, $j=1,...,\tau$, $A_j$ is the number of apps in $\Phi_j$\tabularnewline
\hline
 $K_a$& App profile consisting of apps $a_{i,j}$ and their categories $\Phi_j$ \tabularnewline
\hline
 $\mathcal{G}$& Set of interests in Google interests list\tabularnewline
\hline
\multirow{2}{*}{$\Psi_l$} &  Interest category in $\mathcal{G}$, $l=1,...,\epsilon$, $\epsilon$ is the number of interest \tabularnewline
 & categories defined by Google \tabularnewline
\hline
 $S_g$& Subset of Google interests in $\mathcal{G}$ derived by $S_a$\tabularnewline
\hline
 $I_g$& Interest profile consisting of ${g_{k,l}}$, ${g_{k,l}} \in S_g$\tabularnewline
\hline
 ${g_{k,l}}$& An interest in $I_g$, $k \in G_l$, $l \in \epsilon$ \tabularnewline
\hline
 $S_g^{i,j}$& Set of interests derived by an app $a_{i,j}$ \tabularnewline
\hline
 $D$& The set of demographics defined by analytics companies \tabularnewline
\hline
 ${{d_{m,n}}}$& Demographics options ${d_{m,n}} \in D$, $m=1,...,M$ and $n=1,...,N$ \tabularnewline
\hline
$S_d$& Subset of user's demographics derived by \textit{System App}\tabularnewline
\hline
%\multirow{2}{*}{${d_{o,p,q}}$} &  An ad in the set of ads ${D_{o,p}}$, $o$ is the user profile $o \in \left[ {1..P} \right]$,\tabularnewline
% & $P$ is the number of profiles, $p \in \left[ {1..\tau} \right]$, $q$ is the numbered ads \tabularnewline
%\hline
%\multirow{2}{*}{$f$} & Mapping that returns either direct mapping of the ads category \tabularnewline
% & in Alexa or the ad's keywords ${\kappa_{o,p,q}}$ from Alexa \tabularnewline
%\hline
% ${E_{o,p}}$& Set of ads' with their Alexa transformed categories \tabularnewline
%\hline
% $c$& A specific node in tree-structure of Alexa categories $C$\tabularnewline
%\hline
% $\delta _p^{ran}$& Set of \textit{Random} ads\tabularnewline
%\hline
% $\delta _p^{tar}$& Set of \textit{Targeted} ads\tabularnewline
%\hline
% $\delta _p^{con}$& Set of \textit{Contextual} ads\tabularnewline
%\hline
% $\delta _p^{gen}$& Set of \textit{Generic} ads\tabularnewline
%\hline
\end{tabular}} %}
 \end{center}\caption{List of Notations. \label{table:notations}}
\end{table*}

%\paragraph{\textbf{Demographics in Profile}}
In addition, \textit{ads targeting} is based on demographics so as to reach a specific set of potential customers that are likely to be within a specific age range, gender etc., Google\footnote{Demographic Targeting \url{https://support.google.com/google-ads/answer/2580383?hl=en}} presents a detailed set of various demographic \textit{targeting} options for ads display, search campaigns etc. Hence, profiling also characterises users by various demographics under different options. The demographics $D$ are usually grouped into different categories, with specific options ${d_{m,n}}$, $m=1,...,M$ and $n=1,...,N$, where $M$ is the targeted demographics ranges, e.g. `18-24', `25-34', `35-44', `45-54', `55-64', `65 or more', and `Male', `Female', `Rather not say', and $N$ is the number of different demographic \textit{targeting} options e.g. household income, parental status, location etc. Note that for simplicity, we do not consider the user history and leave it for future work. We suggest that the profiling is done on the user side (i.e. done by the \textit{System App}\footnote{The \textit{System App} is similar to the AdMob SDK \cite{ullah2020protecting} that communicates with Google Analytics for deriving user profiles.}, see Figure \ref{ad-system-blockchain}).

Following sub-section details derivation of \textit{Interests  profile}, which we call the profile \textit{Establishment} process.

\subsection{Mapping Rules for Profiling}\label{subsection:mapping-rules}
We now deliberate the profiling rules as a result of utilisation of mobile apps. The advertising networks derive \textit{Interest profiles} for wide range of audiences based on the collected information and accordingly \textit{target} the mobile users. Note that the mapping rules outlined in this section are based on our observations \cite{ullah2020protecting} \cite{ullah2014characterising} \cite{tchen2014}.
%that consists of mapping of \textit{Apps profile} to \textit{Interests profile}

\paragraph{\textbf{Profile Establishment}} User profile is established by using the installed apps for certain amount of time, after it has sufficient interactions with analytics services. We assume that there is a mapping of \textit{Apps profile} to an \textit{Interests profile} defined by Google Analytics, which is used by analytics companies to individually characterise user's interests across the advertising ecosystem. A sample mapping is also given in \cite{ullah2020protecting} where we observed mapping of \textit{Apps profile} to \textit{Interests profile} by a series of various \textit{profiling} experiments. This mapping includes the conversion of apps categories $\Phi _j$ to interest categories $\Psi_l$. This mapping converts an app ${a_{i,j}} \in {S_a}$ to interest set $S_g^{i,j}$ after a specific level of activity ${t_{est}}$. The ${t_{est}}$ is the  \textit{establishment threshold} i.e. time an app should be used in order to establish profile's interests. The set $S_g^{i,j}$ derived by an app ${a_{i,j}}$ can be represented as a set of interests ${{g_{k,l}}}$ along with their corresponding categories ${{\Psi _l}}$:

%\begin{equation}
%M\left( {{a_{i,j}},{t_{est}}} \right) = \left\{ {\left\{ {{a_{i,j}} \to S_g^{i,j}} \right\}:{a_{i,j}} \in {K_a}} \right\}
%\end{equation}

\begin{equation}
%S_g^{i,j} = \left\{ {\left\{ {{g_{k,l}},{\Psi _l}} \right\}:{g_{k,l}} \in \mathcal{G}} \right\}
S_g^{i,j} \subset \left\{ {\left\{ {{g_{k,l}},{\Psi _l}} \right\}:{g_{k,l}} \in G} \right\}
\end{equation}

We note that all installed apps, from a specific mobile device, do not necessarily contribute to the profile\footnote{The full list of installed apps on an Android based phone is included in the library.db file \url{/data/data/com.android.vending/databases/library.db}}, in which case the $S_g^{i,j}$ is an empty set.

Finally, the resulting \textit{Interests profile} $I_g$ is established as the combination of individual interests derived by each app. The \textit{Interests profile} can be expressed as the union of all the interests sets $S_g^{i,j}$ derived by individual app ${a_{i,j}}$ in ${S_a}$; hence Eq. \ref{eq:initial-interes-profile} can be re-written as follows:

\begin{equation}\label{eq:final-interes-profile}
%{I_g} = \left\{ {\left\{ {\bigcup\limits_{\forall {a_{i,j}} \in \,{K_a}} {S_g^{i,j}} :{g_{k,l}},{\Psi _l}} \right\}:{g_{k,l}} \in G} \right\}
%{I_g} = \left\{ {\bigcup\limits_{\forall \left( {i,j} \right)\,:\,{a_{i,j}} \in \,{K_a}} {S_g^{i,j}} } \right\}
{I_g} = \bigcup\limits_{\forall \left( {i,j} \right)\,:\,{a_{i,j}} \in \,{S_a}} {S_g^{i,j}}
\end{equation}

Furthermore, \textit{System App} derives users' demographics $S_d$ from their settings (e.g. Gmail settings) or by inferring their demographics based on their activities.
%Google my also store an identifier in the web browser using a ``cookie'' where, based on the visited sites, the browser might be associated with certain demographic categories.
We suggest that the users may `edit' their demographics within the \textit{System App}, i.e. similar to that by visiting Google Ad Settings\footnote{Google Ad personalisation \url{https://adssettings.google.com/authenticated}}.

\begin{equation}\label{eq:demographics}
{I_g} = \bigcup\limits_{\forall \left( {i,j} \right):{a_{i,j}} \in {S_a}} {s_g^{i,j}}  \cup \bigcup\limits_{\forall \left( {m,n} \right):m,n \in {S_d}} {{d_{m,n}}}
\end{equation}

\paragraph{\textbf{Profile Evolution}} The profile is updated, and hence the \textit{ads targeting}, each time variations in the users' behaviour are observed; such as for a mobile user using apps that would map to interests other than the existing set of interests. Let a user uses new set of apps $S{'_a}$, which has no overlap with the existing set of apps $S_a$ that has created $I_g$ i.e., $S'_a \subset {\mathcal{A}}\setminus {S_a}$.

%\begin{equation}
%S{'_a} = \left\{ {\left\{ {a{'_{i,j}}} \right\}:a{'_{i,j}} \in A,a{'_{i,j}} \notin {K_a}} \right\}
%\end{equation}

The newly added set of apps $S{'_a}$ is converted to interests ${g_{k,l}}$ with ${t_{evo}}$ as \textit{evolution threshold} i.e. the time required to evolve profile's interests. Thus, an app ${a{'_{i,j}}}$ in $S{'_a}$ should be used for ${t_{evo}}$ time in order to evolve profile's interests $S_g^{'i,j}$. The sets of interests $S_g^{'i,j}$ derived by $a{'_{i,j}}$ are represented as the union of all interests $I{'_g}$, i.e. $I{'_g} = \bigcup\limits_{\forall \left( {i,j} \right)\,:\,{a_{i,j}} \in \,S{'_a}} {S_g^{'i,j}}$ along with the updated demographics (if any) \(\bigcup\limits_{\forall \left( {m,n} \right):m,n \in {S_d}} {d{'_{m,n}}}\). Hence, the final \textit{Interests profile}, $I_g^f$, after the profile \textit{Evolution} process, is the combination of older interests derived during the profile \textit{Establishment} $I_g$ and during when the profile evolves $I{'_g}$ and is given as:

\begin{equation}
%I_g^f = \left\{ {\left\{ {\bigcup\limits_{\scriptstyle{g_{k,l}} \in {I_g}\hfill\atop
%\scriptstyle g{'_{k,l}} \in I{'_g}\hfill} {g_{k,l}^f,{\Psi _l}} } \right\}:g_{k,l}^f \in G} \right\}
I_g^f = {I_g} \cup I{'_g}
\end{equation}

\paragraph{\textbf{Profile Development Process}} We now elaborate insights of profile \textit{Establishment} and \textit{Evolution} processes performed by Google analytics and adopt the same strategy of profile \textit{Development} process. In order for the \textit{Apps profile} to \textit{establish} an \textit{Interests profile}, a minimum level of activity of the installed apps is required. Furthermore, in order to generate one or more interests, an app needs to have the AdMob SDK. This was verified by testing a total of 1200 apps selected from a subset of 12 categories, for a duration of 8 days, among which 1143 apps resulted the \textit{Interest profiles} on all test phones indicating ``Unknown'' interests. We also note that the \textit{Apps profile} deterministically derives an \textit{Interests profile} i.e., a specific app constantly derives identical set of interests after certain level of activity. We further note that the level of activity of installed apps be within a minimum of 24h period (using our extensive experimentations; we note that this much time is required by Google Analytics in order to determine ones' interests), with a minimum of, from our experimentations,  $24/n$ hours of activity of $n$ apps. For a sophisticated profiling, a user might want to install and use a good number of apps that would represent one's interests. After the 24h period, the profile becomes \textit{Stable} and further activity of the same apps does not result in any further changes.

% , from our experimentations,

\begin{figure}[h]
\begin{center}
\includegraphics[scale=0.24]{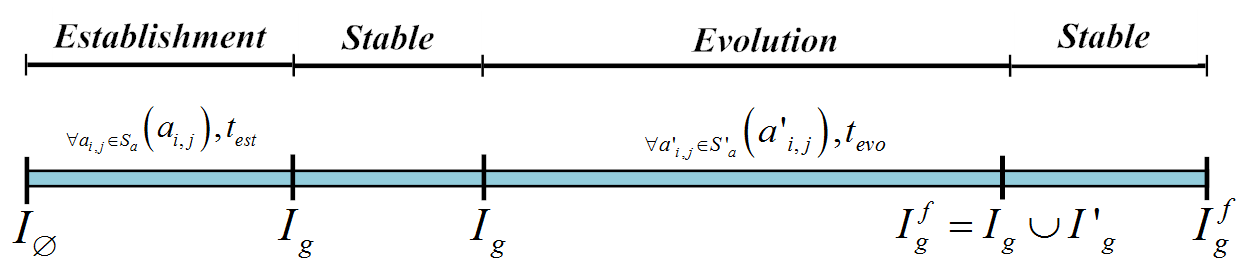}
\caption{Profile \textit{Establishment}  \& \textit{Evolution} processes. ${I_\emptyset }$ is the empty profile before apps utilisation. During \textit{Stable} states, \textit{Interest profiles} $I_g$ or $I_g^f$ remains the same and further activities of the same apps have no effect over the profiles.}
\label{figure-profileEvoEstb}
\end{center}
\end{figure}

Similarly, during the profile \textit{Evolution} process, the \textit{Interests profile} starts changing by adding new interests; once apps other than the existing set of apps $S_a$ are utilised. However, instead of 24h of period of evolving a profile, we observe that the \textit{Evolution} process adds additional interests in the following 72 hours of period, after which the aggregated profile i.e. $I_g^f$ becomes \textit{Stable}. In order to verify the stability of the aggregated profile, we run these apps on 4th day; henceforth we observe no further changes.
%We manually check the profile on each of the phones in 6h intervals (note this cannot be automated).
The mapping of \textit{Apps profile} to \textit{Interests profile} during the \textit{Establishment} and during the \textit{Evolution} process along with their corresponding \textit{Stable} states are shown in Figure \ref{figure-profileEvoEstb}.

\subsection{Problem Statement}\label{problem-statement}
The problem addressed in this paper is where advertising systems profile and track mobile users for their activities and to tailor ads to their specific interests. In complex mobile environment, as opposed to browser-based \textit{ads targeting} where users are tracked and profiled against their visited websites and ad clicks, the users' interests are based on various norms: selection of mobile apps installed and used, (traditional) browser based profiling, ad clicks of certain categories of ads, demographics e.g. age-rank, location etc.; to create a unified user profile. User profiling and its related \textit{ads targeting} expose sensitive information about users \cite{ullah2014profileguard, ullah2020protecting}, e.g. the target could browse through gambling websites or spends excessive time using gambling apps, revealing (including a third-party, such as the website owner) to the advertising systems that the user is a heavy gambler.

In particular, we address two privacy attacks: \textit{Legitimate user profiling} by Analytics, in traditional advertising systems, these functionalities are implemented via an analytics SDK \cite{han2012study} for reporting user's activities to Analytics \& Ads networks, hence, observe various requests to/from mobile users. \textit{Indirect privacy attack}, involves third parties that could intercept and infer user profiles based on \textit{targeted} ads, since ad's traffic is sent in clear text \cite{tchen2014}. We presume that users do not wish to be exposed to their private interests and are willing to receive relevant ads based on their interests.

In addition, another problem being addressed where (user) clients or servers outsource their functionalities (e.g. a firewall to monitor incoming/outgoing transactions for authorising various entities to access particular service/resource) to a third party e.g. to cloud system (\texttt{CS}). Hence, in addition to private \textit{ads targeting}, the user wishes to have private outsource functionalities where any third party, including \texttt{CS}, does not learn the \textit{policies} implemented. %An important thing is to decouple both (mobile) clients and (advertising) servers for their interactions, hence, various functionality (i.e. policies) are implemented by cloud on their (both client's and server's) behalf e.g. tracking user's for their activities, requesting for ads, \textit{billing} for \textit{ad presentation} and \textit{ad click} etc.

\subsection{Threat Model}
Our privacy goal in this work is to decouple, and hence limit the information leakage to either advertising systems or third party, (mobile) users from (advertising) servers for their interactions, hence, various functionality (i.e. \textit{policies}) are implemented by cloud on their (both user's and server's) behalf. In addition, we limit the information leakage for various other activities e.g., tracking user's for their activities, \textit{billing} for \textit{ad presentation} and \textit{ad click} etc. To an extent, we emulate the ``traditional'' setting where the mobile user's profiling is entirely implemented on user's side i.e. \textit{Miner}, hence revealing no information of user profiling to advertising systems. The \textit{Miner} requests for corresponding encrypted ads associated with hash of profile's interests. Hence, the \texttt{CS} only learns the match for some unknown (hashed profile's interests) data along with the corresponding other (encrypted ads) data pointers, i.e. any entity in \texttt{CS} is assumed to be an \textit{honest-but-curious}.

%In the next section, we discuss the experimental setup so as to analyse the resultant profiles during the \textit{Establishment} and \textit{Evolution} processes. We further elaborate the experimentations systemised for collection of served ads in various apps categories (note that we collect all these ads during the \textit{Stable} state) and further demonstrate the procedure for ads collection process.

%\subsection{Effect of Adding New Apps/Establishment Process}

%% file: system-model.tex
\begin{figure*}[h]
\begin{center}
\includegraphics[scale=0.42]{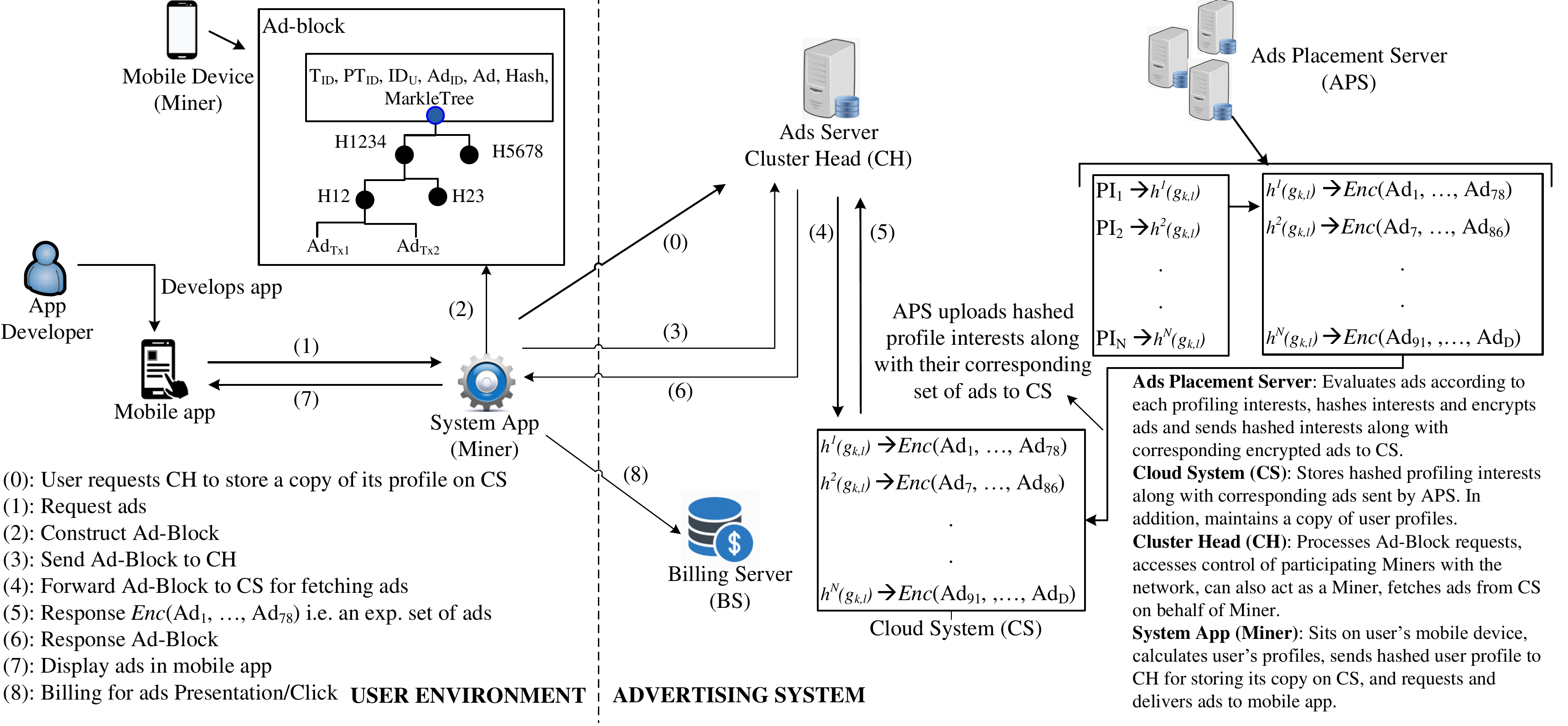}
\caption{A framework for secure user profiling and Blockchain-based \textit{targeted} advertising system for \textit{in-app} mobile ads.}
\label{ad-system-blockchain}
\end{center}
\end{figure*}

Figure \ref{ad-system-blockchain} shows various components of proposed system.

\subsection{Evaluating Advertisements} The Ad Placement Server (\texttt{APS}) groups advertisements according to various interest categories for \textit{targeting} users with specific interests. The advertising system calculates as a similarity match between individual interests and the set of ads; where individual interests and ads are characterised by a set of keywords\footnote{The advertising agency and the advertiser agree on various keywords so as to describe advertisements.}. The similarity is calculated based on the $tf \cdot idf$ metric \cite{manning2008introduction}. We note that this strategy is not metric specific and that other similarity metrics might be used. Let $D$ is the set of ads from all the advertisers wanting to advertise with an advertising agency. Each ad $ A{d_{ID}}$ is represented with a unique identifier $ID=1,..., D$ and is characterised by $\kappa _{A{d_{ID}}}$ keywords. Let the function $f$ returns the set of keywords of an ad i.e. $f\left( {A{d_{ID}}} \right) = {\kappa _{A{d_{ID}}}}$ from the pre-defined list of ads and their corresponding keywords. These set of keywords are then used by another function $g$ for calculating an appropriate interest(s) by calculating the similarity between the ad's keywords \(\kappa _{A{d_{ID}}}\) and the list of keywords of individual interests i.e. \(\kappa _{k,l}\).

%\begin{equation}
%g\left( {{\kappa _{A{d_{ID}}}}} \right) = {\kappa _{k,l}}:sim\left( {{\kappa _{k,l}},{\kappa _{A{d_{ID}}}}} \right) = \bigcup\limits_{\forall Ad_{ID}^' \in D} {\max \left( {simsim\left( {{\kappa _{k,l}},{\kappa _{Ad_{ID}^'}}} \right)} \right)}
%\end{equation}

\[g\left( {{\kappa _{A{d_{ID}}}}} \right) = {\kappa _{k,l}}:sim\left( {{\kappa _{k,l}},{\kappa _{A{d_{ID}}}}} \right) = \]
\begin{equation}
\bigcup\limits_{\forall Ad{'_{ID}} \in D} {\max \left( {sim\left( {{\kappa _{k,l}},{\kappa _{Ad{'_{ID}}}}} \right)} \right)}
\end{equation}

This similarity \(sim\left( {{\kappa _{k,l}},{\kappa _{A{d_{ID}}}}} \right)\) based on $tf \cdot idf$ is calculated as follows: Let $t$ be a particular keyword in \({{\kappa _{A{d_{ID}}}}}\) and $d$ be the set of keywords in \({\kappa _{k,l}}\); then the \textit{term frequency} (occurrences of  a particular keyword $t$ in $d$) is $tf_{t,d}$. The \textit{inverse document frequency} of $t \in {\kappa _{A{d_{ID}}}}$ within $d$ is calculated as $idf_t= log \frac {N}{df_t}$, where $N$ is the collection of keywords of interests in \textit{Interest} categories and the $df_{t}$ is the \textit{document frequency} that is the number of \textit{Interest} categories that contain $t$. The $tf \cdot idf$ is then calculated as $tf \cdot id{f_{t,d}} = t{f_{t,d}} \times id{f_t}$. Thus the score for ${\kappa _{A{d_{ID}}}}$ can be calculated as:

\begin{equation}
score\left( {{\kappa _{A{d_{ID}}}},d} \right) = \sum\limits_{t \in {\kappa _{A{d_{ID}}}}} {tf \cdot id{f_{t,d}}}
\end{equation}

The similarity \textit{score} ranges \(0 \le score \le 1\) for individual ads, where an ad (or a set of ads) with highest similarity match to profile interests is calculated as a candidate ad for $ g_{k,l}$ and is assigned to particular interest e.g. \({g_{k,l}} \leftarrow A{d_{57}}\). Note that a single ad can be assigned to various profiling interests, hence, targeting vast majority of audience. This process is continued for entire set of advertisements. Note \cite{ullah2014characterising} that the served ads are either URLs of the sponsoring merchants (advertisers) or they are related to e.g. the Google Play Store apps/games\footnote{\url{https://play.google.com/store/apps}} i.e. \textit{self-advertising}. Recall, an app marketplace includes a set of $\mathcal{A}$ mobile apps/games \({a_{i,j}}\), grouped in different categories, characterised by a set of keywords i.e. \({{\kappa _{i,j}}}\). We argue that such advertisements can either be evaluated through the \(sim\left( {{\kappa _{k,l}},{\kappa _{i,j}}} \right)\) similarity metric or a direct match between the profiling \textit{Interests'} $\Psi_l$ and \textit{Apps'} $\Phi _j$ categories e.g. using Jaccard Index \(\frac{{{\Psi_l}}}{{{\Phi_j}}}\).

\subsection{Hashing Ads and Encrypted Profile Interests} All the advertisements are associated with various profile interests, as mentioned in previous step. The advertising server now evaluates the hashes for all the interests \({g_{k,l}}:\forall k \in {G_l},l \in \epsilon \) under individual categories \({\Psi _l}\), i.e. \(h\left( {{g_{k,l}}} \right)\); in addition, the corresponding ads are encrypted (e.g. using \texttt{RSA}) i.e. \(Enc\left( {A{d_{ID}}} \right)\), we call this dataset a `\texttt{profiling-ads}'. Note that this operation is done just once during system initialisation or every time new ads are added to the system for advertising. Each hash references various set of ads e.g. \(h\left( {{g_{k,l}}} \right) \to Enc\left(A{d_{ID1}}, \ldots ,A{d_{IDN}}\right)\), which means that a user with specific \(h\left( {{g_{k,l}}} \right)\) will be \textit{targeted} with \(A{d_{ID1}}, \ldots ,A{d_{IDN}}\) ads.

\subsection{Storing \texttt{profiling-ads} in \texttt{CS}} Once the set of profiling interests and their corresponding ads (i.e. \texttt{profiling-ads}) are ready, the \texttt{APS} can upload it to Cloud Storage (\texttt{CS}), which is carried out through an anonymous process similar to those discussed in \cite{dorri2016blockchain}. Note that \texttt{APS} need to go through the \textit{access policy} mechanism to upload (store) the profiling interests and corresponding ads. For this, \texttt{APS} sends a request to \texttt{CS} for storing \texttt{profiling-ads}, requests the pointer to the first block \({{P_b}}\), and the confirmation for availability of storage space. The \texttt{CS} evaluates \textit{access policy} for \texttt{APS} to confirm whether the request is coming from legitimate party and to allow \texttt{APS} to use storage space. Once the \texttt{APS} is allowed to store the \texttt{profiling-ads} data; the \texttt{CS} encrypts the \({{P_b}}\) with $PK+$ and sends it to \texttt{APS} along with the the \texttt{storage-block} $B$ i.e. the capacity of storage block within \texttt{CS} for saving information on physical disk for \textit{spanned} organisation of storing data. Following, the \texttt{APS} evaluates the blocking factor $bfr$ to calculate the number of block to be occupied by \texttt{profiling-ads} and sends it to the \texttt{CS}. Subsequent, \texttt{APS} decrypts the \({{P_b}}\) with $PK-$ to extract the \({{P_b}}\), creates a random ID, and starts sending \texttt{profiling-ads} to \texttt{CS}. The \texttt{CS} calculates the hash of the received data, compares it with the received hash, and (upon its validity) stores the \texttt{profiling-ads} data. Since \texttt{APS} needs to send (a high volume) data in several transactions; for this, the \texttt{CS} sends a new \({{P_b}}\) (encrypted with the $PK+$) to \texttt{APS}.

\subsection{Miner (System App)} \label{miner}
The \textit{System App} sits on a user's devices, which performs a range of various tasks: deriving user profiles as outlined in Section \ref{subsection:mapping-rules}, uploads hashed user profiles to \texttt{CS}, requests ads from \texttt{CS} through \texttt{CH}, and delivers requested ads to mobile apps while fulfilling advertising quota, discussed in detail in Section \ref{ad-block-structure}. The \textit{Miner} processes all incoming and outgoing requests using a range of various transaction types, as discussed in Section \ref{transaction-types}. The \textit{Miner} authorises various apps participating in profiling or requesting ads via \textit{access policy} (discussed in Section \ref{access-policy}) and authenticates various transaction types. In addition, \textit{Miner} stores the list $PK+$ allocated to each mobile app, which the \textit{Miner} only uses to evaluate the request coming from a legitimate app (see Section \ref{access-policy} for details). Note that, as opposed to legacy approach towards ad request or click \cite{tchen2014, ullah2014profileguard, ullah2017enabling} i.e. encapsulating app' name; only the app's $PK+$ is encapsulated in \textit{ad-block} that ensures anonymity and privacy. \textit{Miner}'s additional functions include: evaluating \textit{billing} for ads, generating \textit{Genesis} transaction, distributing and updating keys, forming \textit{ad-block}, monitoring user's activity for apps, updating and uploading user profiles on removal of mobile apps.

Note that, in order to calculate user profiles, the \textit{Miner} communicates, similar way as AdMob SDK\footnote{\url{https://admob.google.com/}} interacts, with installed mobile apps. We are planning an enhanced version of \textit{System App} for future work aiming at user profiling (e.g. frequency of apps usage) for optimising user targeting.

\subsection{Billing}\label{ad-system-billing}
\sloppy We suggest the use of mining any Cryptocurrencies for our \textit{billing} mechanism. We already proposed a \textit{billing} mechanism within an advertising environment \cite{ullah2017enabling}, using Priced Symmetric Private Information Retrieval (PSPIR) \cite{henry2011practical} that is originally proposed for private ecommerce along with polynomial commitments\footnote{A polynomial commitment \cite{kate2010constant} allows constant-sized commitments to polynomials (independent of their degree), where the committed value is an evaluation of this polynomial for specific inputs; consequently, only this value (rather than the polynomial coefficients) will be revealed in the opening phase.} and Zero Knowledge Proofs (ZKP)\footnote{ZKP \cite{damgaard1998commitment} is an interactive protocol that enables one party, the prover, to prove to the verifier that a particular statement is true, while the latter learns no additional information regardless of any a priori knowledge he may posses.}. However, we suggest the use of Cryptocurrency for privacy, secure payment and for compatibility with the underlying proposed advertising system over Blockchain.

To enable private \textit{billing}, we denote the price tags \(C_{prs}^{A{d_{ID}}}\) and \(C_{clk}^{A{d_{ID}}}\) respectively for \textit{ad presentation} and \textit{ad click} operations. The \textit{Advertiser} \(I{D_{ID}}\) buys the \textit{airtime} from ad agencies for advertising their ads. To initiate \texttt{Billing} transaction for buying \textit{airtime}; the advertiser uses her \textit{private key} ($PK-$) and signs message with the amount of Cryptocurrency (e.g. Bitcoin) and \texttt{Billing} server's address, requesting a transaction. The \textit{billing} request transaction is bind with other transactions and is broadcasted to the network, where another \textit{miner} validates the transaction. The process of transaction completes and the \texttt{Billing} server receives its portion of Cryptocurrency in her \textit{wallet}. Similarly, the \textit{Miner} initiate \texttt{Billing} transaction for ads \textit{presentations}\footnote{For reducing processing and communication overhead and enhanced operation, the \textit{Miner} initiates \texttt{Billing} transaction once sufficient ads are shown to the user within a session whereas the \texttt{Billing} transaction for \textit{click} operation is initiated as soon a \textit{click} operation is triggered.} or \textit{clicks} respectively by encoding the \(C_{prs}^{A{d_{ID}}}\) and \(C_{clk}^{A{d_{ID}}}\) price tags within \texttt{Billing} transaction. Note that the \textit{Miner} keeps track of ads presented to a particular app in \textit{tracking-list}, which is forwarded to \texttt{CH} only for those ads that were not presented until the next (e.g. 24hrs) session for the purpose of informing \texttt{CS} about the fulfillment of advertising quota. We use the following notations for representing various symbols/entities: price tags \(C_{prs}^{A{d_{ID}}}\) and \(C_{clk}^{A{d_{ID}}}\) for ad \textit{presentation} and \textit{click}; various \textit{wallets} i.e. \textit{App Developer}'s \(walle{t_{I{D_{APP}}}}\), \textit{Advertiser}'s \(walle{t_{A{D_{ID}}}}\), \texttt{Billing} Server's \(walle{t_{BS}}\); and (Bitcoin) addresses, i.e. \(Ad{d_{I{D_{APP}}}},Ad{d_{A{D_{ID}}}},Ad{d_{BS}}\). Future work is planned for integrating complete implementation and applicability of Cryptocurrency in \textit{billing} for advertising system.

\subsection{Access Control}\label{access-policy}
The \textit{Miner} implements \textit{access policy} (i.e. \textit{local-policy}), holding an \textit{access control list} that allows apps regarding evaluation of different operations e.g. to check whether an app is allowed to request for ads, \textit{billing} etc. The \textit{policies} are also implemented at \texttt{CS} (i.e. \textit{outsource-policy}, the \texttt{CS} processes \textit{policies} on behalf of \textit{Miner} i.e. executes \textit{policy} on incoming transaction requests destined for accessing various resources i.e. accessing ads) to allow access to various ads evaluated for various profile interests and to \texttt{CH} for processing \textit{ad-block} for requesting ads from \texttt{CS} on \textit{Miner}'s behalf. Note that, instead of introducing a \textit{policy header} (as suggested by \cite{dorri2016blockchain}), we evaluate the access from various options present in the \textit{ad-block} (as shown in Figure \ref{block-structure}) for efficient access and reduced overhead.

\begin{figure}[h]
\begin{center}
\includegraphics[scale=0.45]{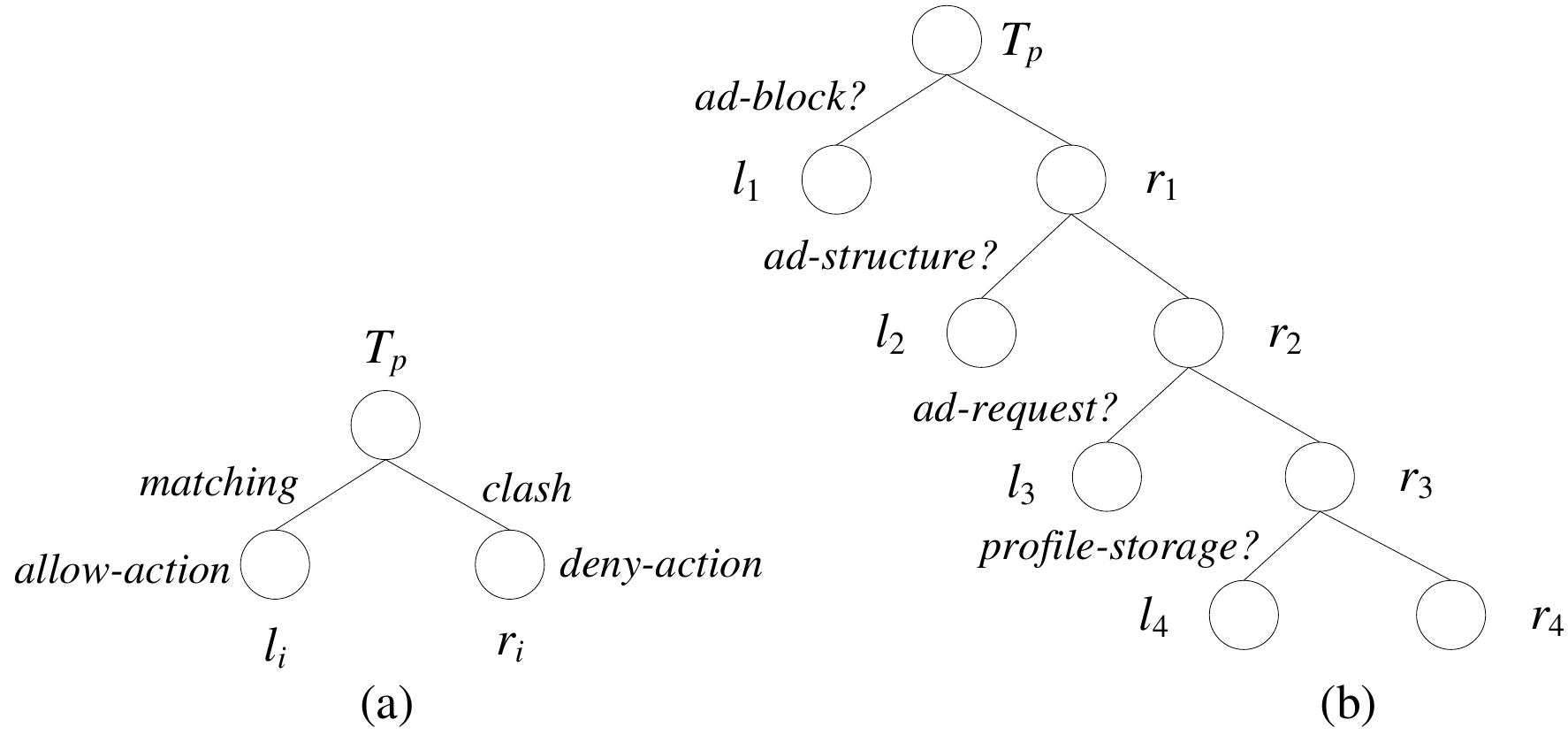}
\caption{An example implementation of \textit{policy tree} along with various \textit{actions} (a) and a simple \textit{access policy} implemented at \texttt{CS} for \textit{Miner} (b).}
\label{access-policy}
\end{center}
\end{figure}

As shown in Figure \ref{block-structure}, we represent the \textit{policies} as binary tree $T_p$ consisting of \textit{left child} \({l_i}\left( {{T_p}} \right)\) and \textit{right child} \({r_i}\left( {{T_p}} \right)\) along with the \textit{matching function} \(m\left( {{l_i}\left( {{T_p}} \right)} \right)\), which is represented as an edge between different (child) nodes and the \textit{action function} \(a\left( {{l_i}\left( {{T_p}} \right)} \right)\). The pair \(\left( {m,a} \right)\) is a \textit{Policy}. There is a list of \textit{action functions} $A$ (mainly allow or disallow and route to next level of $T_p$) and the \textit{matching functions} $M$ i.e. the exact \textit{policy} to implement, such that $a \in A$, $m \in M$. An example set of $M$ that \textit{Miner} implements are: Calculate ads quota, Apps to participate in \textit{billing}, \textit{Profile establishment}, Show ads within app, Apps to change \textit{public key} etc. A simple example \textit{policy} structure is given in Figure \ref{access-policy} (b) implemented at \texttt{CS} that consists of four pairs of distinct \textit{policies}: \(\left( {{m_1},{a_1}} \right),\left( {{m_2},{a_2}} \right),\left( {{m_3},{a_3}} \right),\left( {{m_4},{a_4}} \right)\). The \texttt{CS} use these \textit{policies} to permit \textit{Miner} for requesting various \textit{actions}. Here, $m_1$ is a \textit{match} for evaluating \textit{Miner} for its permission for requesting \textit{ad-block}, likewise, $m_4$ for storing profile in \texttt{CS}.

\sloppy We suggest to use Multi-version of \textit{access policy} i.e. creating a separate copy for \(read\left( {{T_p}} \right)\) and \(modify\left( {{T_p}} \right)\) \textit{policy document}. This will ensure that the \(read\left( {{T_p}} \right)\) copy is only accessed for reading purpose, hence, a number of transactions are allowed to access the copy of \textit{access policy}. Note that the \textit{Miner} has to indicate the type of transaction, as mentioned \textit{Trans. Type} in Figure \ref{block-structure}, under the \texttt{Ad-Block-Structure}.

%% file: ad-block-structure.tex
\sloppy In this section, we present the block structure for various transactions for an \textit{ad-request/response} for operating within the Blockchain network.

\subsection{Ad Block Header Request/Response}\label{sec-ad-block-request-response}
\begin{figure*}[h]
\begin{center}
\includegraphics[scale=0.45]{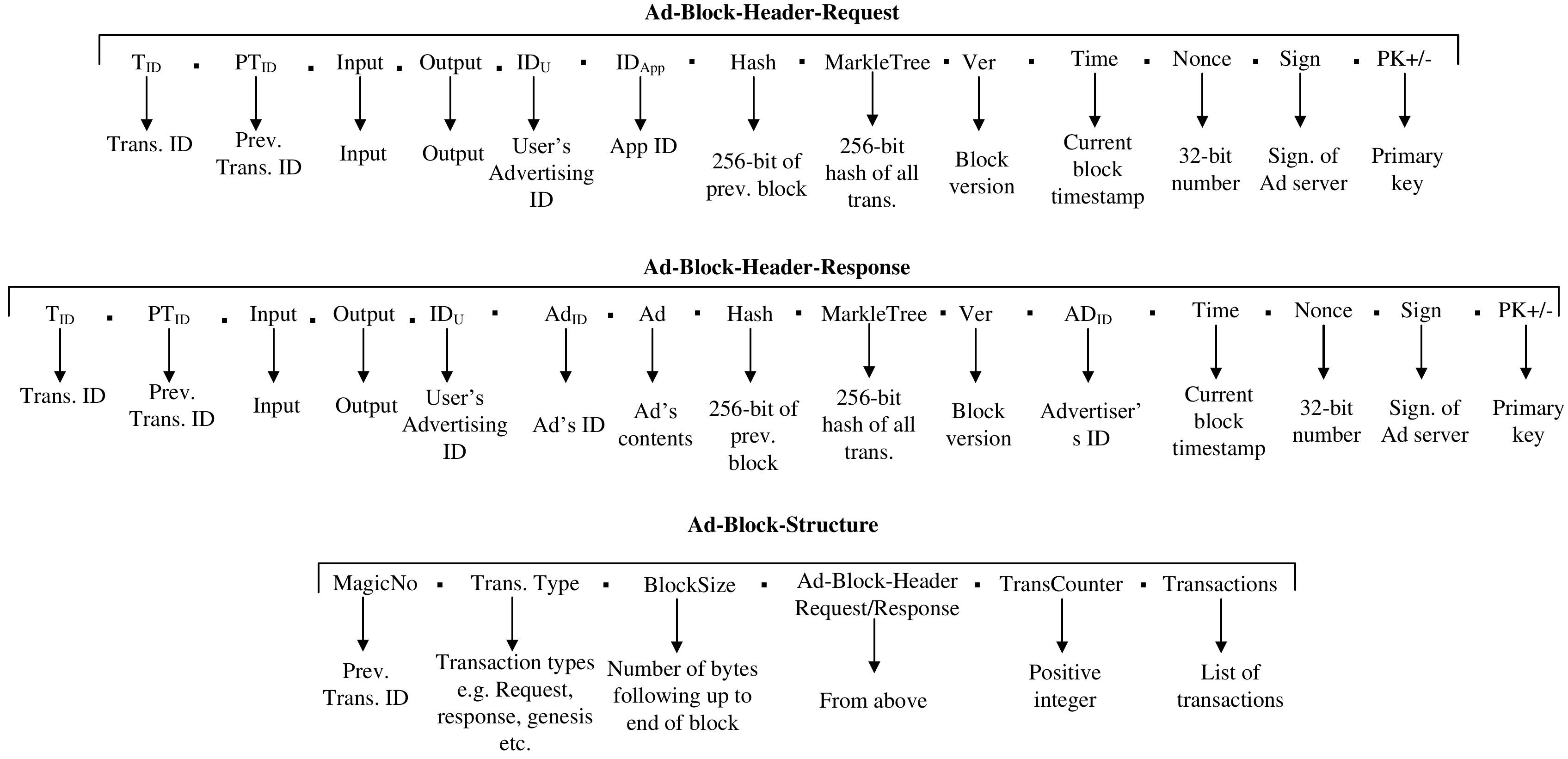}
\caption{Block structure of an `\texttt{Ad-Block-Header-Request}' (upper), `\texttt{Ad-Block-Header-Response}' (middle), and `\texttt{Ad-Block-Structure}' (lower) for an \textit{ad-request/response} block in a Blockchain.}
\label{block-structure}
\end{center}
\end{figure*}

\sloppy Figure \ref{block-structure} presents the structure of a transaction (e.g. an \textit{ad-request}) `\texttt{Ad-Block-Header-Request}' (upper) for \textit{ad-request}, `\texttt{Ad-Block-Header-Response}' (middle) for \textit{ad-response} and the `\texttt{Ad-Block-Structure}' (lower) in a Blockchain network scenario. This block is generated by the \textit{System App}, a \textit{Miner} sitting in user's mobile devices, for requesting an ad (for showing in an app) form \texttt{CH}. The \({T_{ID}}\) denotes the transaction's unique identifier, which is the hashed value of the other fields of this transaction. Note that the first transaction (or block) is called the \textit{Genesis} transaction, which lays the foundation on which additional blocks are sequentially added to form (sequential ledger by connecting current transaction with the previous transaction using \(P{T_{ID}}\)) a chain of blocks. \(P{T_{ID}}\) represents the identifier of the previous transaction that is used for establishing connection between the current and previous transactions. The \textit{input} i.e. all the request ad transactions are made to the \textit{Miner}'s address, enlists all the ads that the \texttt{CH} has already sent to the \textit{Miner} within a particular session. Note that both the \texttt{CH} and \textit{Miner} keep track of the already served ads so that they are not presented again to user within the same session. We take the session as the time duration where user utilises a particular mobile app until she exits the application or there is no activity for a considerable (e.g. 24hrs) time duration. The \textit{Ad-block} with used ``\textit{input}'' is then served to a different (\textit{Miner}) user, which upon used by the current \textit{Miner} becomes and stored in the \textit{output} field.

%\begin{figure*}[h]
%\begin{center}
%\includegraphics[scale=0.5]{figs/block-structure.jpg}
%\caption{Structure of an `Ad-Block-Header' (upper) and `Ad-Block-Structure' (lower) for an ad request.}
%\label{block-structure}
%\end{center}
%\end{figure*}

The \(I{D_U}\)\footnote{\(I{D_U}\) can be reset via the Google Settings System App on Android devices} represents (user) advertising ID that identifies user's mobile device (detailed discussion over \(I{D_U}\) can be found in \cite{ullah2017enabling}). We suggest that \(I{D_U}\) does not change for a particular session. This will help in keeping track of ads shown to the users, in addition to, \textit{billing} where app developer's (\(I{D_{App}}\)) \textit{wallet} is populated with particular amount of Cryptocurrency. Note that for simplicity we represent \(I{D_{App}}\) both for identifying app's developer (Developer ID) and name of the mobile app. The \(A{d_{ID}}\) is the ID of the advertisement shown to the user. The (mobile) \textit{Miner} also keeps track of particular ads in order to fulfil ad's advertising quota\footnote{This requirement comes from the advertising agency, where a particular is presented for certain frequency \cite{ullah2014characterising, nath2015madscope}.}; Table \ref{ads-quota} presents an example tracking of advertising quota e.g. the advertisement `DEF' still needs to be presented 50\% app airtime.

\begin{table}[h]
%\begin{center} \scalebox{0.5}{
\begin{center} \scalebox{0.80}{
{
\begin{tabular}{|c|c|r|r|}
\hline
Date/Time & Advertisement & Frequency & Served\tabularnewline
\hline
04/02/2020:10:45:52 & ABC & 100 & 65\%\tabularnewline
\hline
05/02/2020:01:23:41 & DEF & 67 & 50\%\tabularnewline
\hline
\end{tabular}
}} \caption{An example tracking advertising quota.\label{ads-quota}}
 \end{center}
\end{table}

Furthermore, the \(Ad\) is the actual content of the advertisements that are presented to users within mobile apps; the contents of an ad varies over different sizes of \textit{Min} ad size of $12KB$, the \textit{Max} size of $20KB$ and the \textit{Avg} ad of $16KB$, as observed in \cite{ullah2017enabling}. Similarly, \(A{D_{ID}}\) is the advertiser's ID of particular advertisement(s), which is used to deduct advertiser's Bitcoin \textit{wallet} (detailed discussion over \textit{billing} is given in \ref{ad-system-billing}).

\sloppy To enable communication in a Blockchain network, the peers of a Blockchain network require a \textit{Public-Private Key} pair i.e. represented with $PK+/-$ in Figure \ref{block-structure}, which is used to increase the anonymity. The \textit{Sign} field contains the signature of the transaction generator (e.g. \textit{Miner} (\textit{System app})) using the \textit{public-private key} pair. Following illustrates to understand the exchange of data between \textit{Miner} and \texttt{CH} (e.g., \textit{Miner} sends an \textit{ad-request} to \texttt{CH}): \textit{Miner} encrypts the \textit{ad-request} transaction data using \texttt{CH}'s \textit{public key} $PK+$ and then creates a signature by taking the hash of \textit{ad-request} and encrypts it using her (\textit{Miner}'s) \textit{private key} $PK-$. The transaction data is encrypted along with the digital signature is included in the \textit{Ad-Block-Header-Request/Response}. Subsequent, this transaction is broadcasted over the Blockchain network, since it is addressed\footnote{The peer's address in a Blockchain network is obtained by calculating the cryptographic hash of user's \textit{public key} e.g. in Bitcoin, the \texttt{SHA-256} encryption algorithm is used for deriving user addresses \cite{nakamoto2019bitcoin}.} to \texttt{CH}, hence \texttt{CH} starts verifying the transaction contents by decrypting the digital signature using \textit{Miner}'s \textit{public key} while decrypts the transaction data using her (\texttt{CH}'s) \textit{private key}. Note that \texttt{CH} verifies the transaction data by comparing the \textit{ad- request} hash to the hashed data in digital signature, detailed discussion can be found in \cite{zheng2018blockchain}. We suggest to, similar to \cite{dorri2019mof}, store the \textit{public key} in $PK+/-$ field for reducing the size of the transaction and for future proof transactions i.e. to prevent from malicious attacks where intruders could possibly reconstruct \textit{private key} using \textit{public key}.

%Recall that a Miner orders the pending transactions into a block until it reaches the pre-defined block size, as mentioned, blocks in a Blockchain are connected through hashes of previous blocks. The transactions, included in block, are being hashed by the Miner as part of the Merkle tree; the transactions stored are lead to the Merkle root stored in the block header. As mentioned in \cite{nakamoto2019bitcoin}, the block header being part of the longest chain with Merkle root can be used for different purposes, such as Simple Payment Verification (SPV), and it also speeds up the process of verifying the membership of a transaction in a block. An example Merkle tree, for 8 different Ad requests (leaf nodes represented with \(A{d_{TX1}}\) to \(A{d_{TX8}}\) on the left), including internal hashed nodes and Merkle root node is shown Figure \ref{merkle-tree}. The hash value of a transaction (e.g. an ad request) is stored in these leaf nodes, note that we presume a binary Merkle tree that has \({2^{n - 1}}\)  leaf nodes with $n$ height. An example slice of a Merkle tree with $n=8$ is also shown on the right of Figure \ref{merkle-tree}. The yellow nodes are the potential hashed nodes that potentially could be pruned for reclaiming disk space, details for node pruning can be found in \cite{nakamoto2019bitcoin}.

\begin{figure}[h]
\begin{center}
\includegraphics[scale=0.25]{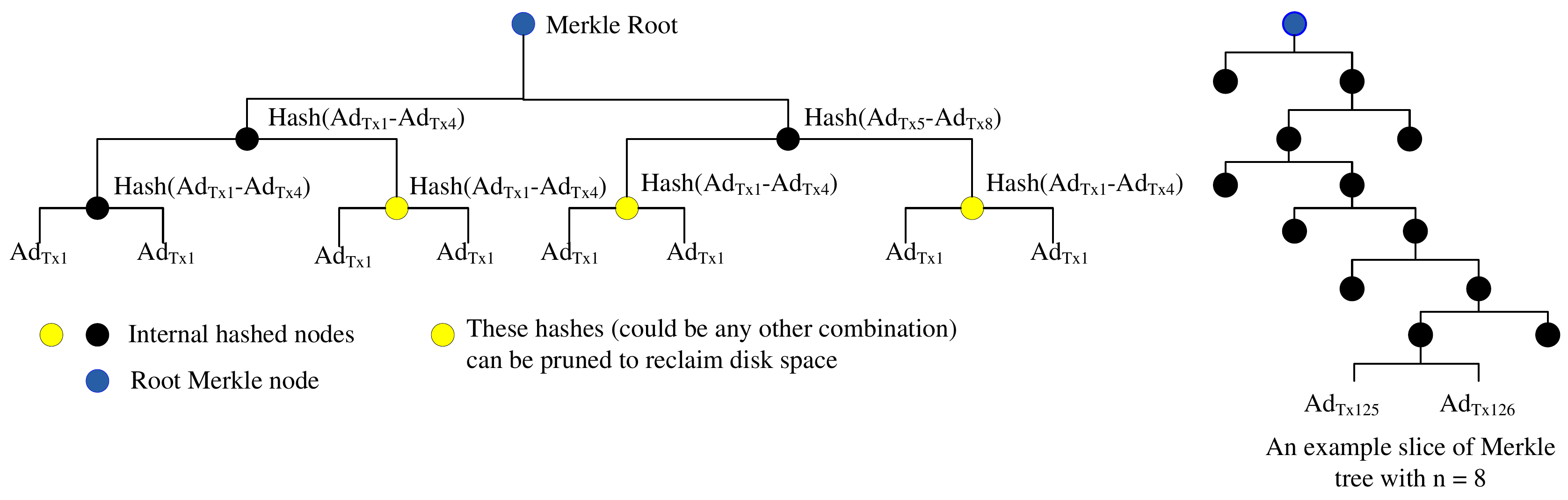}
\caption{Ad requests hashed in a Merkle tree (left). A slice of Merkle tree for a height of $n=8$ (right).}
\label{merkle-tree}
\end{center}
\end{figure}

%\begin{figure*}[h]
%\begin{center}
%\includegraphics[scale=0.5]{figs/merkle-tree.pdf}
%\caption{Ad requests hashed in a Merkle tree (left). A slice of Merkle tree for a height of $n=8$ (right).}
%\label{merkle-tree}
%\end{center}
%\end{figure*}
%
The transactions, included in block, are being hashed by the \textit{Miner} as part of the Merkle tree; the transactions stored are lead to the Merkle root stored in the block header. As mentioned in \cite{nakamoto2019bitcoin}, the block header being part of the longest chain with Merkle root is used for different purposes, such as Simple Payment Verification (SPV), and it also speeds up the process of verifying the membership of a transaction in a block. An example Merkle tree, for 8 different \textit{ad-requests} (leaf nodes represented with \(A{d_{TX1}}\) to \(A{d_{TX8}}\) on the left), including internal hashed nodes and Merkle root node is shown Figure \ref{merkle-tree}. The hash value of a transaction (e.g. an \textit{ad-request}) is stored in these leaf nodes; note that we presume a binary Merkle tree that has \({2^{n - 1}}\)  leaf nodes with $n$ height. An example slice of a Merkle tree with $n=8$ is also shown on the right of Figure \ref{merkle-tree}. The yellow nodes are the potential hashed nodes that potentially could be pruned for reclaiming disk space, details for node pruning can be found in \cite{nakamoto2019bitcoin}.

\subsection{Ad Block Structure}
The lower part of Figure \ref{block-structure} shows the structure of an \textit{Ad-block} `\texttt{Ad-Block-Structure}' consisting of the \texttt{Ad-Block-Header Request/Response} and the list of some or all recent (pending) \textit{ad-request} transactions. Recall that a \textit{Miner} orders the pending transactions into a block until it reaches the pre-defined block size, as mentioned, blocks in a Blockchain are connected through hashes of previous blocks.

\subsection{Ad Block Transaction Types}\label{transaction-types}
Various types of communication done among different parties of a Blockchain are enabled via transaction, where different types of transactions are designated for various purposes. E.g. a \texttt{Billing} transaction is enabled for calculating \textit{billing} for various ads presented in mobile apps or when the user clicks on particular ads \({A{d_{ID}}}\), as presented in Section \ref{ad-system-billing}. Similarly, the \texttt{Access}, \texttt{Upload}, and \texttt{Update} transaction types are respectively used for \textit{accessing} profile\footnote{Note that the users upload profiles to \texttt{CS}, as explained in Algorithm \ref{algo-miner-profiling}, which can be accessed by the user when she signs-in to another device or losses her profile.}, \textit{uploading} the profile, or when the user profile \textit{changes}. In addition, the \texttt{Request} and \texttt{Response} transaction, as discussed above, for requesting ads from \texttt{CS} or response ads from \texttt{CS}. Furthermore, the \texttt{Genesis} transaction is when the \textit{Miner} starts participating in Blockchain network. Note that the \texttt{Genesis} transaction is generated for every installed app or for every new app being installed on the device, however, grouped by the app developer \(I{D_{App}}\). Hence, all the activities i.e. \textit{ad-request}, \textit{ad-response}, \textit{billing} \(I{D_{App}}\) and \({A{D_{ID}}}\) both for presented and clicked ads, and participation in profiling are chained together and are tracked for all these activities from current block to the \texttt{Genesis} block. In addition, a \texttt{Remove} transaction type is used to uninstall an app, indicating that the user might not be interested in this specific category of app and also its contribution to user profile. Lastly, the \texttt{Monitor} transaction is generated by the \textit{Miner} to monitor activity of each app for various operations of \textit{ad-request}, \textit{ad-response} etc. Note that the transactions are secured via asymmetric encryption, digital signatures and cryptographic hash functions (e.g. \texttt{SHA-256}). Furthermore, the lightweight hashing \cite{bogdanov2011spongent} is employed to detect any changes in the transaction's content.

%% file: adblock-operation.tex
This section presents the design requirements of a Blockchain-based mobile advertising system and the design of various algorithms describing comprehensive operations of its various components.

\subsection{Design Requirements}
The system is designed by taking into account various requirements: privacy, reduced overhead over (user) client, compatibility of \textit{billing} with Cryptocurrency, efficiency at ad network and client.

\paragraph*{\textbf{Privacy}} The underlying design uses Blockchain technology for preserving user privacy, hence various transactions are buried under the structure of Blockchain transaction for \textit{ad-request}, \textit{ad-click}, \textit{billing} etc., hence, the ad network does not learn any information from these various operations. In traditional advertising setting, the \texttt{APS} establishes user profiling with the help of cookies \cite{ullah2014characterising} and communicates with the mobile devices for various other operations of ad presentation, click etc. However, in proposed method, the \textit{Miner} constructs user profiles locally on the user device\footnote{Among other functionality, an important feature of the proposed architecture is to reduce interaction between user device and advertising system, hence the ad system will not be able to learn more about user's activities.}, hashes of profile interests \({g_{k,l}}\) and uploads to \texttt{CS} or utilise it for various operations including \textit{billing}. The \texttt{CS} learns no information about user's interests, either during \textit{policy} checking or accomplishing various operations.

\paragraph*{\textbf{Reduced overhead over (user) client}} It is critical to impose marginal overhead over \textit{Miner}, hence, it is made optional for user devices to participate as a \textit{Miner} or request \texttt{CH} to construct \textit{ad-request}, henceforth, the \textit{Miner} sends \({h\left( {I{D_U}} \right)}\) and \({h\left( {I{D_{APP}}} \right)}\). However, it is crucial for \textit{Miner} to evaluate various other essential operations (i.e. a trade-off between privacy and operational overheads) in order to enable a private and secure advertising system, e.g., constructing user profile and uploading to \texttt{CS}, keep updating the \textit{tracking-list}, \textit{billing} for ads, and initiating the \textit{Genesis} transaction. The aim is to achieve reduced infrastructure cost, efficiency, and flexibility.

\paragraph*{\textbf{Compatibility of billing with Cryptocurrency}} The biggest industries, such as travel, banking sectors, are inclined towards adapting the Cryptocurrency that will bring industries, government and community together. In addition, the number of mobile internet users have surpassed 4 billion mark\footnote{\url{https://99firms.com/blog/mobile-marketing-statistics/}} and is set to continue growing in the future. This proposal is an effort to adapt the Cryptocurrency in advertising system for providing a platform for embedding Cryptocurrency for \textit{billing} various parties within the ad system.

\paragraph*{\textbf{Efficiency at ad network and (user) client}} The Blockchain based advertising system should be computationally fast with reduced computation and, in specific, communication cost. As evident in \cite{ullah2014characterising}, in traditional \textit{in-app} advertising setting using Google AdMob, a  mobile app requesting an ad and the triggered action from e.g. a user click, consists of a series of different steps: The mobile app sends a \texttt{POST} message passing various information about the device including phone version, model, app running on phone etc.; the ad server sends ad containing web address, \texttt{JavaScript}, static objects; cookie for tracking user's action; the web server associated with the ad is contacted and the landing page is downloaded and is shown to the user. Note that the mobile app developer agree on integrating ads in mobile apps by embedding ad/analytics library. It was also evident that the ads are refreshed every 15--20sec and all the above steps are followed for every single ad presentation and its associated click, hence, severely affecting throughput. This proposal reduces the communication overhead by sending a pool of ads with single \textit{ad-request} transaction and locally serving ads. In addition, the \texttt{APS} uploads the \texttt{profiling-ads} set only during the system initialisation or when new advertiser is added to the system.

Following illustrates the idea behind various operations of the Blockchain-based advertising system.
\subsection{The Blockchain-based Advertising System}
The initial step is performed by \texttt{APS} via Algorithm \ref{algo-global}. This includes evaluating \texttt{profiling-ads} and then uploading the profile interests \({{g_{k,l}}}\) and corresponding ads \({A{d_{ID}}}\) to \texttt{CS}. %An important thing to note is that the \texttt{APS} applies the

\begin{algorithm}[h]
\SetKwInOut{Input}{Input}
\SetAlgoLined
 \caption{\texttt{Global Setup (\texttt{APS})}} \label{algo-global}
 \Input{\({{g_{k,l}}}\); \({A{d_{ID}}}\); \({{L_{k,l}}}\)}
 Run \({{L_{k,l}}}\) = \texttt{Profiling} (\({g_{k,l}}\); \({A{d_{ID}}}\))\\
 Run \texttt{Storage} \({\left( {{L_{k,l}}} \right)}\)\\
\end{algorithm}

%\begin{algorithm}[h]
%\SetAlgoLined
% \caption{\texttt{Global Setup (APS)}} \label{algo-global}
% \KwInput{\({{g_{k,l}}}\); \({A{d_{ID}}}\); \({{L_{k,l}}}\)}
% Run \({{L_{k,l}}}\) = \texttt{Profiling} (\({g_{k,l}}\); \({A{d_{ID}}}\))\\
% Run \texttt{Storage} \({\left( {{L_{k,l}}} \right)}\)\\
%\end{algorithm}

Once the \texttt{APS} completes evaluating \texttt{profiling-ads}, it uploads this dataset to \texttt{CS} using Algorithm \ref{algo-profile-storage} requesting the pointer to the first block $P_b$ for data storage. We assume that the storage in \texttt{CS} consists of various blocks of size $B$ i.e. \texttt{storage-block}, where each $B$ can accommodate numerous records, hence to evaluate various data records that could be fit in one block. The \texttt{APS} sends the \textit{bfr} i.e. Blocking factor, to \texttt{CS} to accordingly reserve the storage space for \texttt{profiling-ads}. Following, the \texttt{APS} starts sending the \texttt{profiling-ads} and progressively asks the $P_b$ to next available $B$.

%Following algo is for evaluating ads relevant to user profiles.

\begin{algorithm}[h]
\SetKwInOut{Input}{Input}
\SetAlgoLined
 \caption{\texttt{Profiling(\({g_{k,l}}\); \({A{d_{ID}}}\))}} \label{algo-profiling}
 \Input{\({g_{k,l}}\); \({A{d_{ID}}}\)}
%\KwInput{\({g_{k,l}}\); \({A{d_{ID}}}\)}
 Initialise empty list \({L_{k,l}}\) with $k \times l$ cells.\\
 \For{\(ID = 1, \ldots ,D\)}{
    \({\kappa _{A{d_{ID}}}} \leftarrow f\left( {A{d_{ID}}} \right)\)\\
    \For{$l=1,..., \epsilon$}{
        \For{$k=1,...,G_l$}{
            \sloppy Evaluate: \(core\left( {{\kappa _{A{d_{ID}}}},d} \right) = \sum\limits_{t \in {\kappa _{A{d_{ID}}}}} {tf.id{f_{t,d}}} \); \(\forall A{d_{ID}},\forall {g_{k,l}}\)
        }
        \(\bigcup\limits_{\forall Ad{_{ID}} \in D} {\max \left( {sim\left( {{\kappa _{k,l}},{\kappa _{Ad{'_{ID}}}}} \right)} \right)} \)\;
        \({L_{k,l}} \cdot add\left( {{g_{k,l}},A{d_{ID}}} \right)\)
    }
 }
 Evaluate hashes of profile interests and corresponding ads in \({L_{k,l}}\).\\
 \For{\(m = 1, \ldots ,len\left( {{L_{k,l}}} \right)\)}{
%    \({h^m}\left( {{g_{k,l}}} \right);{h^m}\left( {A{d_{ID}}} \right)\)\\  % hashed-ads
    \({h^m}\left( {{g_{k,l}}} \right);\) \({En{c_{PK + }}\left( {A{d_{ID}}} \right)}\)\\  % encrypted-ads
%    \(L_{k,l}^* \cdot add\left( {{h^m}\left( {{g_{k,l}}} \right),{h^m}\left( {A{d_{ID}}} \right)} \right)\)   % hashed-ads
    \(L_{_{k,l}}^*.add\left( {{h^m}\left( {{g_{k,l}}} \right),En{c_{PK + }}\left( {A{d_{ID}}} \right)} \right)\)\\   % encrypted-ads
%    \({h^m}\left( {{g_{k,l}}} \right) \cdot add\left( {{h^m}\left( {A{d_{ID}}} \right)} \right)\)
 }
 \texttt{Return} \(\left( {L_{k,l}^*} \right)\)
\end{algorithm}

The \texttt{CS} runs the Algorithm \ref{algo-access-policy} for evaluating access permission. To initiate this process, the \texttt{CS} creates a \(read\left( {{T_p}} \right)\), sets $s$ to the initial node in $T_p$, and starts traversing through $T_p$. Recall an example tree-structured \textit{policy} shown in Figure \ref{access-policy}, representing $k=4$ distinct policies and corresponding \textit{matching criteria} $m$ and their resultant \textit{action functions} $a$. During the traversal, if a \textit{match} is found, the corresponding \textit{action} is performed, however, the \textit{policy traversal} terminates whenever a \textit{leaf} node is entered.

%Following algo is for storing profile-ads on CS.

\begin{algorithm}[h]
\SetKwInOut{Input}{Input}
\SetAlgoLined
 \caption{\texttt{Storage \(\left( {{L_{k,l}}} \right)\)}, \texttt{APS}} \label{algo-profile-storage}
 \Input{\({{L_{k,l}}}\)}
%\KwInput{\({{L_{k,l}}}\)}
 \texttt{APS} requests \texttt{CS} to store \({{L_{k,l}}}\)\\
 \(APS.request\left( {{P_b}} \right)\) from \texttt{CS}\\
  \If{(Run \texttt{Access Policy (\texttt{APS})})}{
  \If{!(storage full)}{
    \texttt{CS} sends  \texttt{storage-block} $B$ to \texttt{APS}.\\
    \(CS.send\left( {En{c_{PK + }}\left( {{P_b}} \right)} \right)\) to \texttt{APS}
   }
   }
 \texttt{APS} sends \(bfr = \left\lceil {{\raise0.7ex\hbox{${len\left( {{L_{k,l}}} \right)}$} \!\mathord{\left/
 {\vphantom {{len\left( {{L_{k,l}}} \right)} B}}\right.\kern-\nulldelimiterspace}
\!\lower0.7ex\hbox{$B$}}} \right\rceil \)\\
 \texttt{APS} \(De{c_{PK - }}\left( {{P_b}} \right)\)\\
 \texttt{APS} generates random ID\\
 \For{\(i = 1, \ldots ,bfr\)}{
 \(APS.send\left( {ID,{L_{k,l}}\left[ i \right],h\left( {{L_{k,l}}} \right),En{c_{PK + }}\left( {{P_b}} \right)} \right)\)\\
  \If{\(h\left( {{L_{k,l}}} \right) = CS.calc\left( {h\left( {{L_{k,l}}} \right)} \right)\) AND Transaction is valid}{
    \(CS.store\left( {{L_{k,l}}\left[ i \right]} \right)\)\\
    \(CS.send\left( {En{c_{PK + }}\left( {{P_b} = {P_b} + 1} \right)} \right)\)\\
   }
 }
\end{algorithm}

The Algorithm \ref{algo-miner-profiling} illustrates the procedure for storing user profiles on \texttt{CS}, once constructed as a result of \textit{Profile development process}, as explained in Section \ref{user-profiling}. The \textit{Miner} first evaluates the cryptographic hash, using \texttt{SHA-256}, of individual profile's interests and then uploads to \texttt{CS} using Algorithm \ref{algo-profile-storage}. In addition, the \texttt{CS} also evaluates Algorithm \ref{algo-access-policy} to allow/disallow \textit{Miner} $I{D_U}$.

%Following algo is for storing user profile on CS.
\begin{algorithm}[h]
\SetKwInOut{Input}{Input}
\SetAlgoLined
 \caption{\texttt{Profiling-Storage (Miner)}} \label{algo-miner-profiling}
 \Input{\({a_{i,j}} \in {S_a}\); \({d_{m,n}} \in {S_d}\)}
%\KwInput{\({a_{i,j}} \in {S_a}\); \({d_{m,n}} \in {S_d}\)}
 Establishment: $I_g$ via use of \({a_{i,j}} \in {S_a}\)\\
 Map $S_a$ to \(S_g^{i,j}\); \({I_g} \leftarrow S_g^{i,j}\)\\
 Evaluate \(d_{m,n}\); \({I_g} \leftarrow d_{m,n}\)\\
 Evolution: \(I_g^f \leftarrow {I_g} \cup S{'{_g^{i,j}}} \cup d{'{_{m,n}}}\)\\
 Evaluate hashes of profile interests.\\
 \For{\(m = 1, \ldots ,len\left( {I_g^f} \right))\)}{
    \({h^m}\left( {{g_{k,l}}} \right)\); \({g_{k,l}} \in S_g^{i,j} \in I_g^f\)\\
    \({I_g^*.add\left( {{h^m}\left( {{g_{k,l}}} \right)} \right)}\)
 }
 Run \texttt{Storage\({\left( {I_g^*} \right)}\), Miner\({\left( {I{D_U}} \right)}\)}
\end{algorithm}

The \textit{ad-request} consist of various steps, as outlined in Algorithm \ref{algo-ads-request} and depicted in left side of Figure \ref{ad-system-blockchain} i.e. \textit{User Environment}. The mobile app $I{D_{APP}}$ first requests \textit{Miner} for the ads to display to user; Step (1) of Figure \ref{ad-system-blockchain}. The \textit{Miner} also implements an \textit{Access Policy} to evaluate the ad request coming from \textit{legitimate} app. Consequently, the \textit{Miner} constructs an \textit{Ad-Block} using \texttt{Ad-Block-Header-Request} by encoding $I{D_U}$, $I{D_{APP}}$ etc., Step (2) of Figure \ref{ad-system-blockchain}. Detailed procedure for constructing \texttt{Ad-Block-Header-Request} is also given in Section \ref{sec-ad-block-request-response}. Following, the \textit{Miner} sends \textit{Ad-Block} to \texttt{CS} (Step (3) of Figure \ref{ad-system-blockchain}), which the \texttt{CH} forwards to \texttt{CS} (Step (4) of Figure \ref{ad-system-blockchain}) for fetching ads according to various $g_{k,l}$ of $I{D_U}$. The \textit{ad-response} (Step (5) of Figure \ref{ad-system-blockchain}) is encoded using \texttt{Ad-Block-Header-Response} as shown in Figure \ref{block-structure} (middle), illustrated in Section \ref{sec-ad-block-request-response}. Finally, the \textit{Miner} sends the fetched ads to mobile app for display within active session (i.e. Step (7) of Figure \ref{ad-system-blockchain}). %Note that the \textit{Miner} also keeps track of various ads shown within mobile app to fulfil the advertising quota (an example tracking advertising quota is also given in Table \ref{ads-quota}), detailed description over \textit{tracking-list} is given in Section {sec-ad-block-request-response}.

%Following algorithm is for requesting ads:

\begin{algorithm}[h]
\SetAlgoLined
 \caption{\texttt{Ads-Request (Miner)}} \label{algo-ads-request}
 \({I{D_{APP}}}\) requests ads from \textit{Miner}\\
 Run \texttt{Access Policy \(\left( {{I{D_{APP}}}} \right)\)}\\
 \textit{Miner} constructs \textit{Ad-Block} using \({I{D_U},I{D_{APP}}, \cdots }\)\\
 \textit{Miner} sends \textit{Ad-Block} to \texttt{CS}\\
 Run \texttt{Access Policy \(\left( {{I{D_{U}}}} \right)\)}\\
 \texttt{CH} forwards request to \texttt{CS}\\
 Run \texttt{Access Policy \(\left( CH \right)\)}\\
 Fetch ads according to \(\left( {{I{D_{U}}}} \right)\) profile interests from \texttt{CS}\\
% Run \texttt{Traverse-Storage \(\left( {{I{D_{U}}}} \right)\)}\\
 Send \(h\left( {A{d_1}, \cdots ,A{d_N}} \right)\) to \texttt{CH}\\
 \texttt{CH} sends set of ads to \textit{Miner}\\
 \textit{Miner} presents ads to Mobile app and updates \textit{tracking-list}\\
\end{algorithm}

%Following algorithm shows billing for ads \textit{presentation} and \textit{clicks}:
The \textit{Miner} corresponds to \texttt{BS} for \textit{billing} for \textit{ad presentation} and \textit{ad click} operations via Algorithm \ref{algo-ads-billing}, see Section \ref{ad-system-billing} for detailed description of \textit{billing}. In order to avoid the communication and computation overhead, we propose to periodically evaluate the \textit{billing} (e.g. every 24hrs) for \textit{ad presentation} through calculating the fulfillment of advertising quota. However, for \textit{click} operation, the \textit{billing} mechanism is initiated as soon as a \textit{click} operation is triggered. Note that $u$ and $v$ are the proportions of \textit{billing} amount (in Cryptocurrency) respectively paid to \(walle{t_{I{D_{APP}}}}\) and \(walle{t_{BS}}\).

\begin{algorithm}[h]
\SetKwInOut{Input}{Input}
\SetAlgoLined
 \caption{\texttt{Billing \(\left( {I{D_{App}},A{D_{ID}},A{d_{ID}}} \right)\)}} \label{algo-ads-billing}
\Input{\(C_{prs}^{A{d_{ID}}}\); \(C_{clk}^{A{d_{ID}}}\); \(walle{t_{I{D_{APP}}}}\); \(walle{t_{A{D_{ID}}}}\); \(walle{t_{AS}}\)}
%\KwInput{\(C_{prs}^{A{d_{ID}}}\); \(C_{clk}^{A{d_{ID}}}\); \(walle{t_{I{D_{APP}}}}\); \(walle{t_{A{D_{ID}}}}\); \(walle{t_{AS}}\)}
 \textbf{Ad Presentation}:\\
 Deduct: \({B_{prs}} = walle{t_{A{D_{ID}}}} - C_{prs}^{A{d_{ID}}}\)\\
 \texttt{Billing} transaction to pay \textit{App Developer}:  \(walle{t_{I{D_{APP}}}} \leftarrow \left( {{\raise0.7ex\hbox{${{B_{prs}}}$} \!\mathord{\left/
 {\vphantom {{{B_{prs}}} u}}\right.\kern-\nulldelimiterspace}
\!\lower0.7ex\hbox{$m$}}} \right)\)\\
 \texttt{Billing} transaction to pay \texttt{Billing} server: \(walle{t_{BS}} \leftarrow \left( {{\raise0.7ex\hbox{${{B_{prs}}}$} \!\mathord{\left/
 {\vphantom {{{B_{prs}}} v}}\right.\kern-\nulldelimiterspace}
\!\lower0.7ex\hbox{$n$}}} \right)\)\\
 \textbf{Ad Click}:\\
 Deduct: \({B_{clk}} = walle{t_{A{D_{ID}}}} - C_{clk}^{A{d_{ID}}}\)\\
 \texttt{Billing} transaction to pay \textit{App Developer}:  \(walle{t_{I{D_{APP}}}} \leftarrow \left( {{\raise0.7ex\hbox{${{B_{clk}}}$} \!\mathord{\left/
 {\vphantom {{{B_{clk}}} u}}\right.\kern-\nulldelimiterspace}
\!\lower0.7ex\hbox{$m$}}} \right)\)\\
 \texttt{Billing} transaction to pay \texttt{Billing} server: \(walle{t_{BS}} \leftarrow \left( {{\raise0.7ex\hbox{${{B_{clk}}}$} \!\mathord{\left/
 {\vphantom {{{B_{clk}}} v}}\right.\kern-\nulldelimiterspace}
\!\lower0.7ex\hbox{$n$}}} \right)\)
\end{algorithm}

%Following profile is used for traversing access policy.
\begin{algorithm}[h]
\SetKwInOut{Input}{Input}
\SetAlgoLined
 \caption{\texttt{Access Policy \(\left( Node\right)\)}} \label{algo-access-policy}
\Input{\(read\left( {{T_p}} \right)\), \(modify\left( {{T_p}} \right)\), $T_p$}
%\KwInput{\(read\left( {{T_p}} \right)\), \(modify\left( {{T_p}} \right)\), $T_p$}
\({s = {l_i}\left( {{T_p}} \right)}\)\\
\While{traverse until $s$ is leaf node}{
 \eIf{\({{l_i}\left( {{T_p}} \right)}\) is not leaf}{
 \({{l_i}\left( {{T_p}} \right)} = 1\)\\
  \eIf{access is allowed}{
    Move to the left child.\\
    \({s = {l_i}\left( {{T_p}} \right)}\)\\
    Evaluate \(\left( {{m_i},{a_i}} \right)\)\\
   }{
   Move to the right child.\\
   \(s = {r_i}\left( {{T_p}} \right)\)\\
  }
   }{
     \If{\({{l_i}\left( {{T_p}} \right)}\) is leaf node}{
     Move to the right child.\\
       \({s = {r_i}\left( {{T_p}} \right)}\)\\
        Exit\\
   }
  }
 }
\end{algorithm}

% \For{not at end of this document}{
%  read current\;
%  \eIf{understand}{
%   go to next section\;
%   current section becomes this one\;
%   }{
%   go back to the beginning of current section\;
%  }
% }

%% file: evaluation.tex
We now discuss our experimental evaluation i.e., the processing time for generating \textit{public-private key} ($PK+/-$) pairs with various key sizes and the encryption and decryption time of advertiser's ads with different number of interests along with its effect over selecting a range of \textit{public-private key} sizes. Note that this wide selection of options affect the \textit{ad presentation} time that include the processing time (including \textit{access policy traversing time}, matching profile hashing at \texttt{CS} for fetching relevant ads, the transmission delay (bigger sizes of advertiser's ads are subject to longer transmission delays), which in turn dominates the overall communication delay i.e. the response back from \texttt{CS} to \textit{Miner} for \textit{ad request}. Following, to analyze the scalability of the our proposed framework, we evaluate the processing delays for varying number of traversed rules and corresponding actions, represented with different shapes of \textit{access policy trees}.

\subsection{Experimental setup}\label{experimental-setup}
\sloppy Evaluations were performed on Intel(R) Core(TM) i7-2620M CPU @2.70GHz and 8GB or RAM; we use a \texttt{Python} module \texttt{cryptography.hazmat.primitives.asymmetric} from \texttt{RSA} for generating $PK+/-$ key pairs and for evaluating \texttt{AES} encryption/decryption time of advertiser's ads with different sizes of ad's contents. We generate 10,000 $PK+/-$ key pairs with a range of selection of various key sizes (i.e. $512, 1024, 2048,$ and $4096$ bits) (i.e., we stress-test the proposed framework with continuously generating a total of 40,000 $PK+/-$ key pairs) and evaluate the key pair generation time. Note that the $PK+/-$ key pairs are generated once and are used for various other operations.

Following, we generate 1000 user profiles by randomly selecting up to 20 different profile interests\footnote{Note that apps generate a very limited number of profile interests, as evident in \cite{ullah2020protecting}.} representing users with diverse interests, equally divide them into 10 groups consisting of 100 user profiles (i.e. \({P_1},{P_2}, \ldots ,{P_{10}}\)). Figure \ref{key-user-profiles}(a) shows user profile groups uniformly distributed from 1000 user profiles that consist of various profile interests with 10.53 $Avg$ interests with a $St.Dev$ of 5.88 interests. We calculate the hashes of these profiles using \texttt{SHA1}, \texttt{SHA224}, \texttt{SHA256}, \texttt{SHA384}, and \texttt{SHA512} hashing schemes.

\begin{figure}[h]
\begin{center}
\includegraphics[scale=0.7]{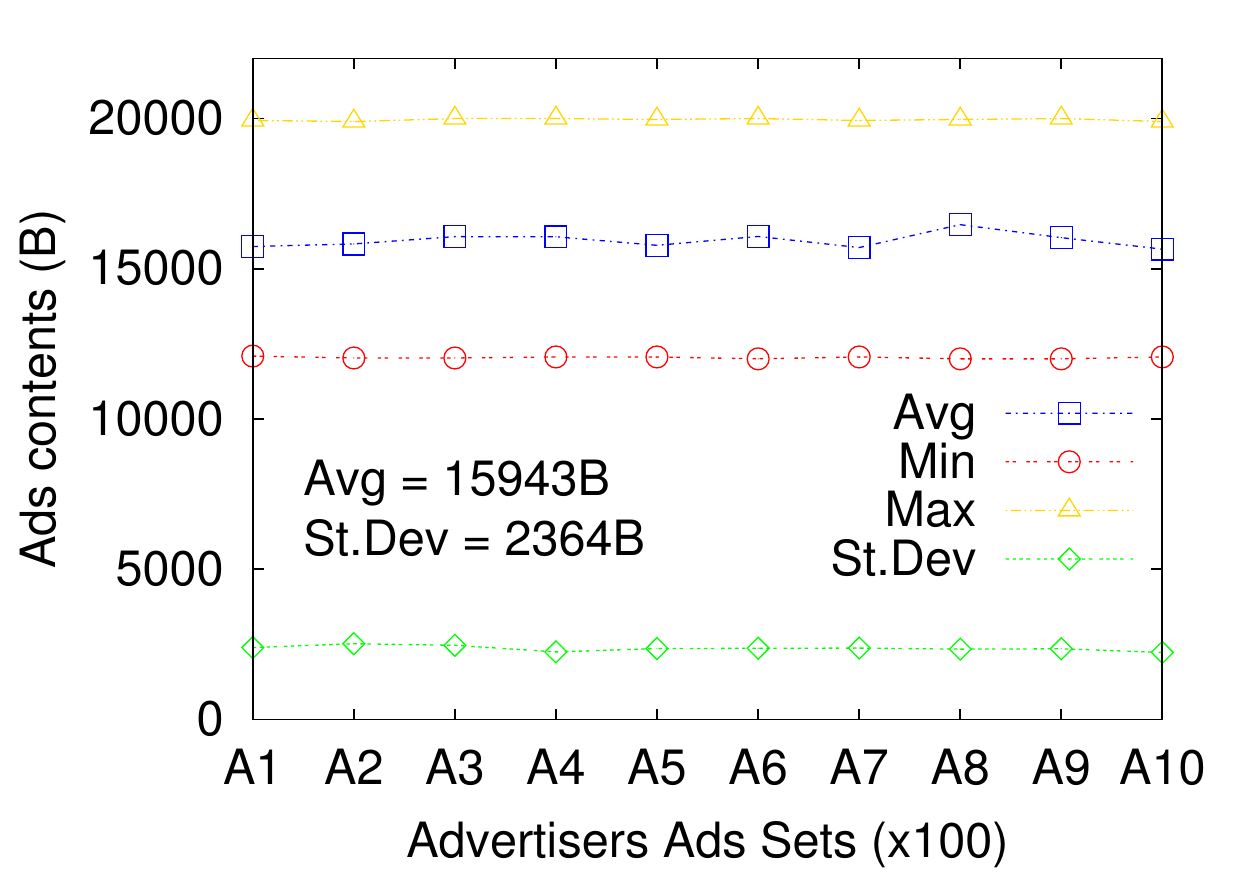}
\caption{Randomly generated ads with $Min=12KB, Max=20KB, Avg=15.94KB, St.Dev=2.364KB$ content's sizes. $A_i=100\,ads$; $i=1, \ldots, 10$.}
\label{ads-distribution}
\end{center}
\end{figure}

Furthermore, we generate 1000 advertiser's ads with random sizes of between $12KB$ to $20KB$\footnote{The selection of these sizes of ads are based on our prior works, which has characterized \textit{in-app} mobile ads \cite{ullah2014characterising, ullah2020protecting}.} and equally divide them into 10 groups consisting of 100 randomly chosen ads (i.e. \({A_1},{A_2}, \ldots ,{A_{10}}\)). Figure \ref{ads-distribution} shows advertiser's ads groups uniformly selected from 1000 ads with various sizes, with an $Avg$ size of 15943B (i.e. 15.94KB) and a $St.Dev$ of 2364B. Following, we evaluate the \texttt{AES} encryption/decryption time of these groups of ads with various key sizes of $512, 1024, 2048,$ and $4096$ bits. Note that in order to encrypt ads of arbitrary length, we use hybrid encryption i.e. \texttt{RSA} to asymmetrically encrypt a symmetric key. To do this, we first randomly generate a symmetric encryption key using \texttt{AES} and encrypt the ads, then, we encrypt the \texttt{AES} key using $PK+$ generated with \texttt{RSA} and transmit both the \texttt{AES}-encrypted ads and \texttt{RSA}-encrypted \texttt{AES} key to the \textit{Miner}. Following, to decrypt the ad's contents, the \textit{Miner} first decrypts the \texttt{RSA} block using $PK+$ to discover the \texttt{AES} symmetric key and then decrypt the symmetrically encrypted ads using \texttt{AES} to discover the resultant set of ads. Similarly, we construct the \texttt{Ad-Block-Structure} for \textit{ad-request/response} by including \texttt{Ad-Block-Header-Request} or \texttt{Ad-Block-Header-Response} (by encoding \({Min=12KB,Max=20KB}\) sizes of ad' content \cite{ullah2017enabling}) respectively for requesting data for \textit{ad presentation} by the \textit{Miner} and the response back from \texttt{CS} to \textit{Miner}.

%\begin{figure}[h]
%\begin{center}
%\includegraphics[scale=0.7]{figs/key-user-profiles.eps}
%\caption{Randomly generated user profile with $Min=1, Max=20, Avg=10.53, St.Dev=5.88$ profile interests. $P_i=100\,user\,profiles$; $i=1, \ldots, 10$.}
%\label{key-user-profiles}
%\end{center}
%\end{figure}

\subsection{Beskpoke Miner}
\begin{figure}[h]
\begin{center}
\includegraphics[scale=0.5]{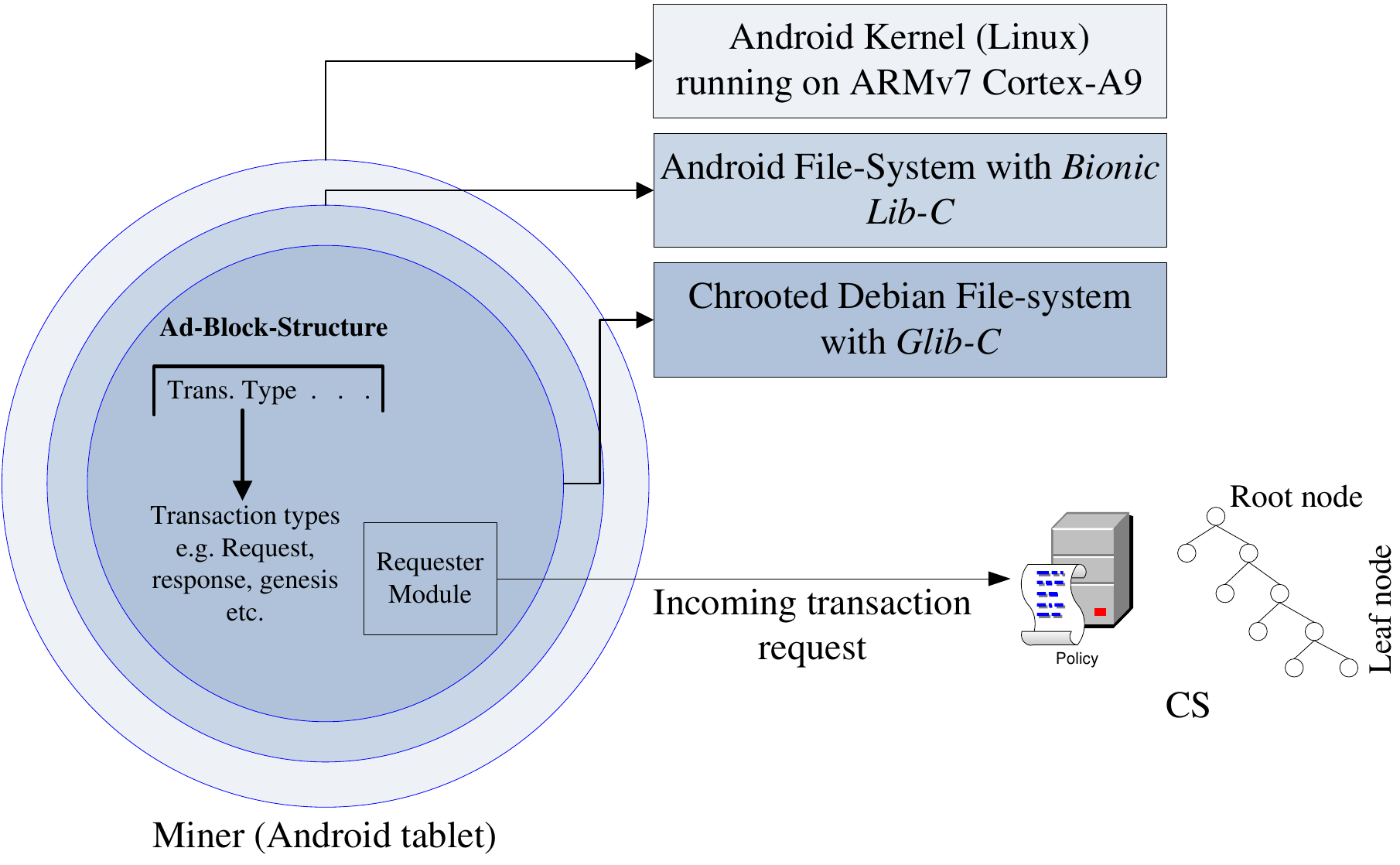}
\caption{Runtime architecture for a bespoke \textit{Miner}.}
\label{bespoke-miner}
\end{center}
\end{figure}

We develop a bespoke version of the \textit{Miner}, we made significant changes to implement some of its functionalities, such as generating an \textit{ad-block}, transactions with different \textit{policy} requests etc., on an Android-based device. We implement \textit{Miner} on Google Nexus 7, ARM Cortex-A9 Nvidia Tegra, 1GB RAM, 1.2 GHz quad-core. To implement various functionalities, first we install CyanogenMod that is a custom ROM for Android devices in order to get \texttt{root access} and \texttt{developer} capabilities. We then add support for the standard Linux utilities, in order to implement various \textit{Miner} functionalities, that are not generally ported as part of the Android OS. The Android OS is equipped with \textit{Bionic} C library\footnote{\url{https://android.googlesource.com/platform/bionic/}} instead of Glibc\footnote{\url{http://www.gnu.org/software/libc/}} i.e. the standard GNU C library, found on most Unix variant operating systems, which enable Android OS to achieve high performing standard C library with small memory and disk footprint. Following, we make both the \textit{Glibc} and \textit{Bionic} coexist on Android platform using the \texttt{chroot} system call in order to use Linux's utilities; this step also enables us to create Ubuntu Linux file system, which we can run a complete Debian environment on top of the Android kernel ARM CPU support OS. These various layers of functionalities are presented in Figure \ref{bespoke-miner}.

To implement a system module of \textit{access policy} and to evaluate the system performance under different shapes of access trees, we set our \texttt{CS} to store $100, 200, \ldots , 1000$ set of rules and matches with root node at \textit{random} positions between $min=1$ to $max=n$, $n=\{100, 200, \ldots , 1000\}$ representing various shapes of trees with nodes (i.e. \textit{matching rules} and corresponding \textit{actions}) present on both sides. E.g. for a tree of 1000 nodes with root node set to 1 or 1000 means that the entire set of \textit{matching rules} and corresponding \textit{actions} are present respectively on the right and left side of the root node, similarly, the set of rules and matches are equally distributed on both sides with the position of root node at position 500. We repeat this experiment for 1000 times for each of various trees of $100, 200, \ldots , 1000$ nodes with \textit{randomly} selecting root nodes at different positions. Note that for each request, in order to \textit{stress test} the system, we deliberately put the \textit{matching rule} for every \textit{access policy} request at the leaf node of any specific tree, hence traversing the entire tree for each matching rules. We configure the \texttt{Requester module} (i.e. inside the \textit{Miner} as depicted in Figure \ref{bespoke-miner}) to launch these requests, with above setting, to \texttt{CS} (we experiment our \texttt{CS} with the same specs as presented in Section \ref{experimental-setup}), as shown at the bottom of Figure \ref{bespoke-miner}. The processing delay is evaluated at \texttt{CS} for all these experiments, along with $Avg,Min,Max,St.Dev$ of root node placement positions and processing delays.

\subsection{Experimental results}
\begin{figure}[h]
\begin{center}
\includegraphics[scale=0.7]{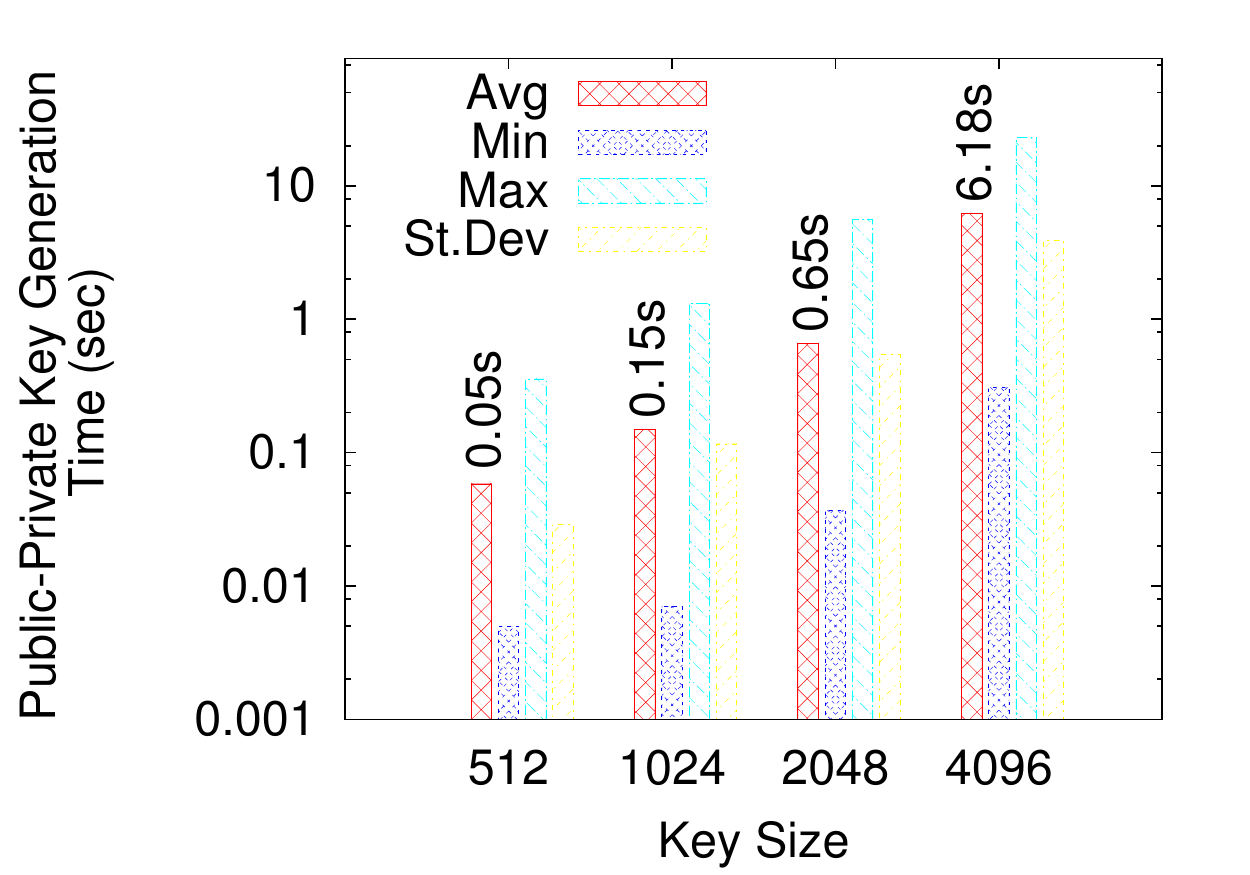}
\caption{Processing delay for generating 40,000 $PK+/-$ key pairs with various key sizes of $512, 1024, 2048,$ and $4096$ bits.}
\label{key-generate-time}
\end{center}
\end{figure}

\paragraph*{\textbf{Processing delay for generating $PK+/-$}}
Our first evaluation is to explore the effect of $PK+/-$ key pair size over the key pair generation time. Figure \ref{key-generate-time} shows the performance in terms of processing delays evaluate to $Avg, Min, Max, St.Dev$ for generating $PK+/-$ key pair with various sizes of $512, 1024, 2048,$ and $4096$ bits. E.g. the $Min$ and $Max$ processing time for generating 10,000 keys with key size of $512$ bits respectively is 0.005sec and 0.35sec with an $Avg$ delay of 0.05sec and 0.03 $St.Dev$. We note that the $Avg$ \textit{power trendline} results in a curve with 0.002sec slope and a 0.86 of $r-squared$ value, as the key size is increased from $512$ to $4096$ bits. The $Avg$ key generation delay is also written on top of each $PK+/-$ key pairs with various key sizes.

\begin{figure*}[h]
     \begin{center}
      \subfigure[User profiles]{\includegraphics[scale=0.6]{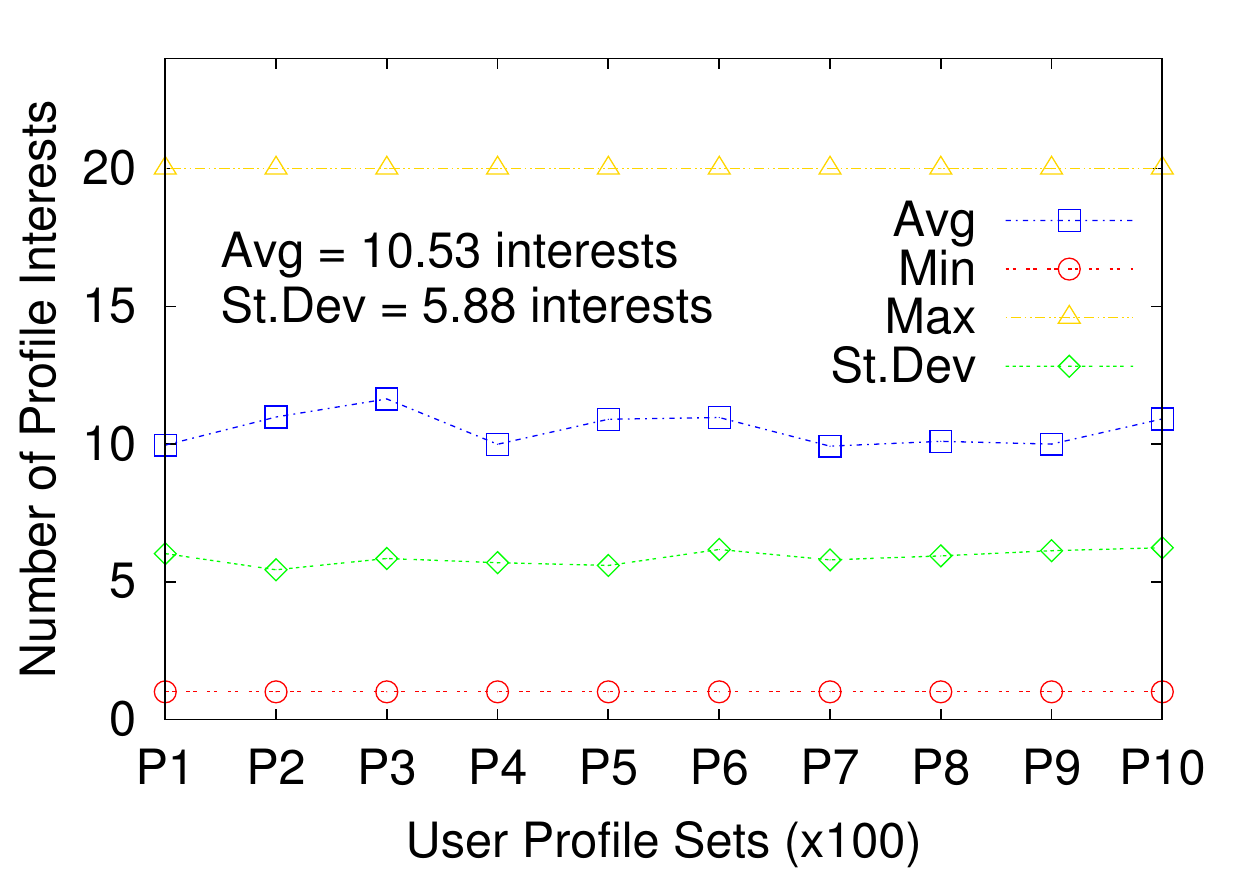}}
      \subfigure[Hashing]{\includegraphics[scale=0.6]{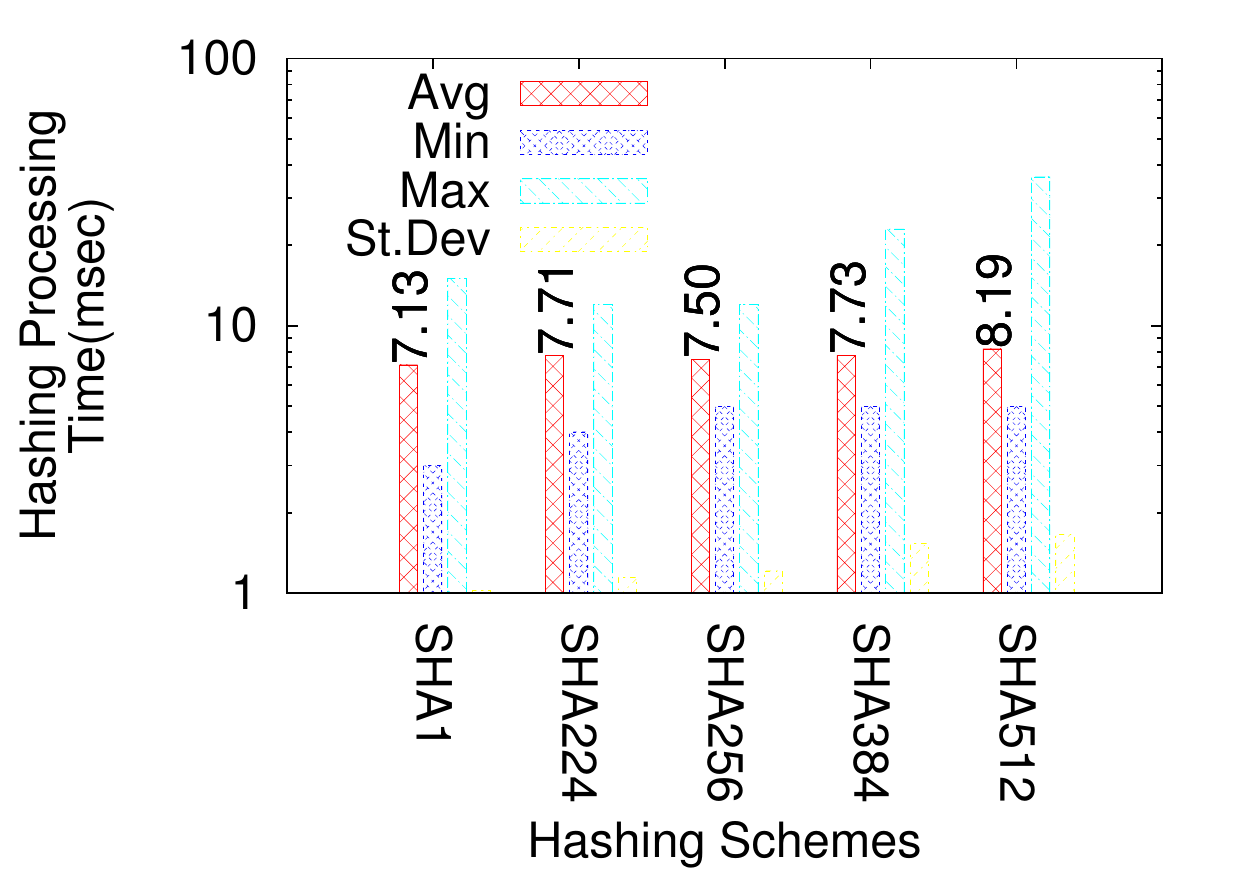}}
     \end{center}
    \caption{Randomly generated user profile (a) with $Min=1, Max=20, Avg=10.53, St.Dev=5.88$ profile interests. $P_i=100\,user\,profiles$; $i=1, \ldots, 10$. Processing time of hashing profiles (b) using \texttt{SHA1}, \texttt{SHA224}, \texttt{SHA256}, \texttt{SHA384}, \texttt{SHA512} hashing schemes.}
     \label{key-user-profiles}
\end{figure*}

\begin{figure*}[h]
     \begin{center}
      \subfigure[1024]{\includegraphics[scale=0.3]{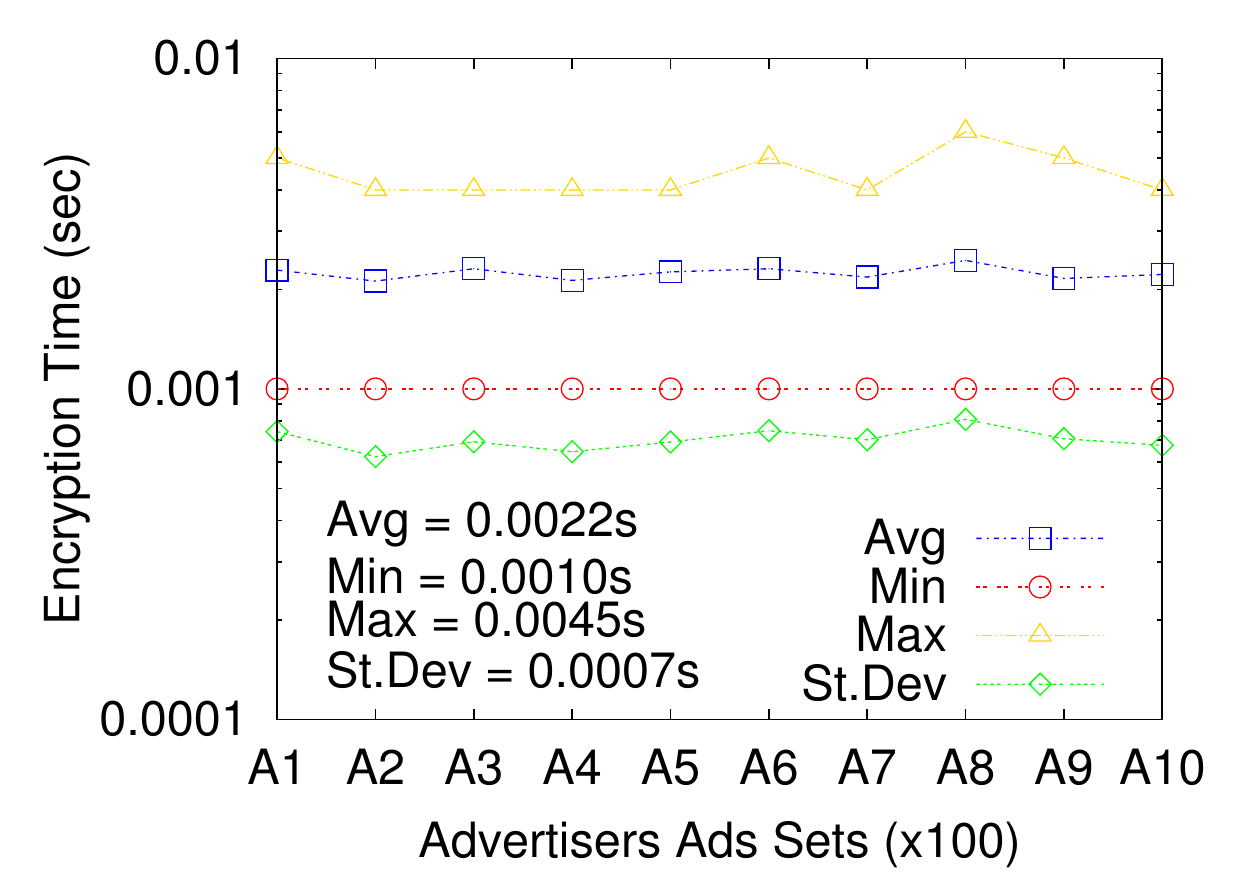}}
      \subfigure[2048]{\includegraphics[scale=0.3]{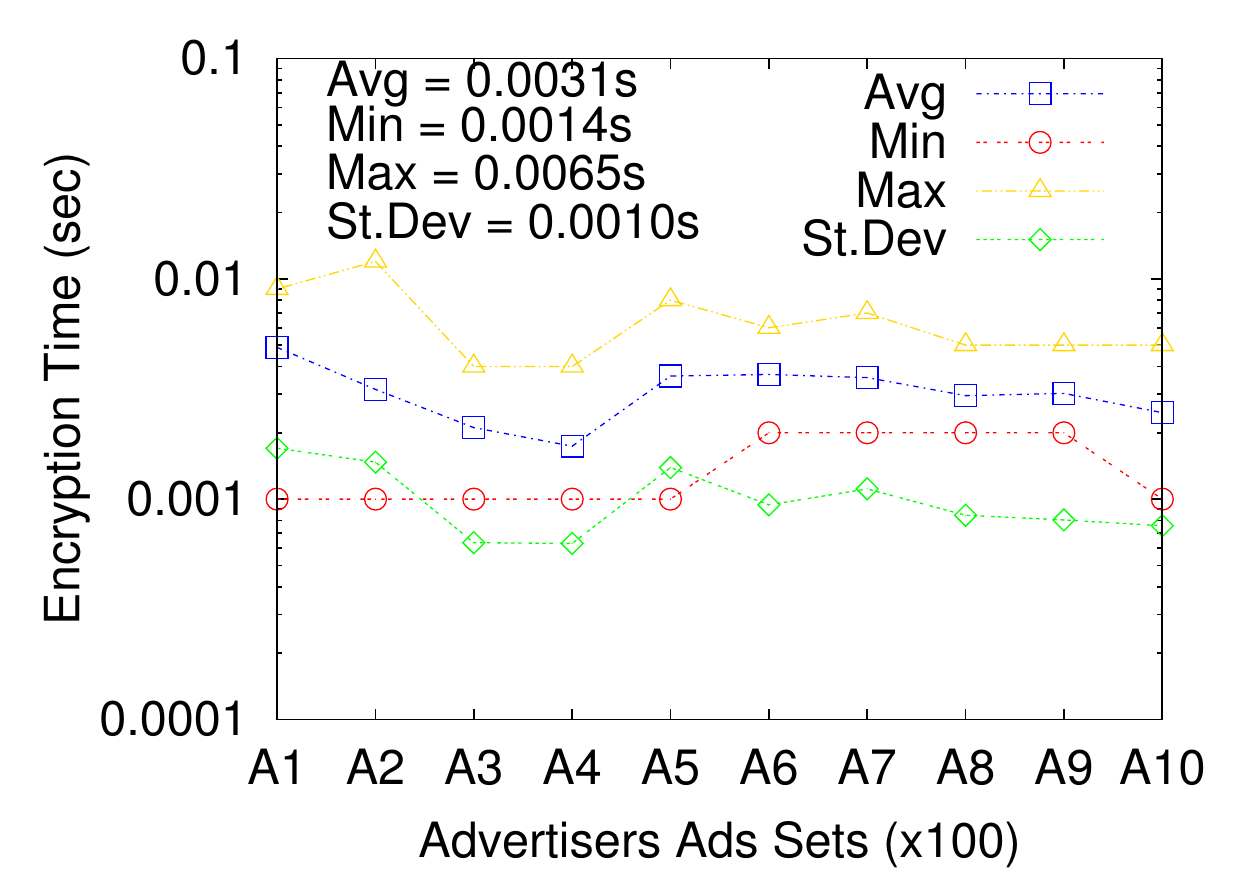}}
      \subfigure[4096]{\includegraphics[scale=0.3]{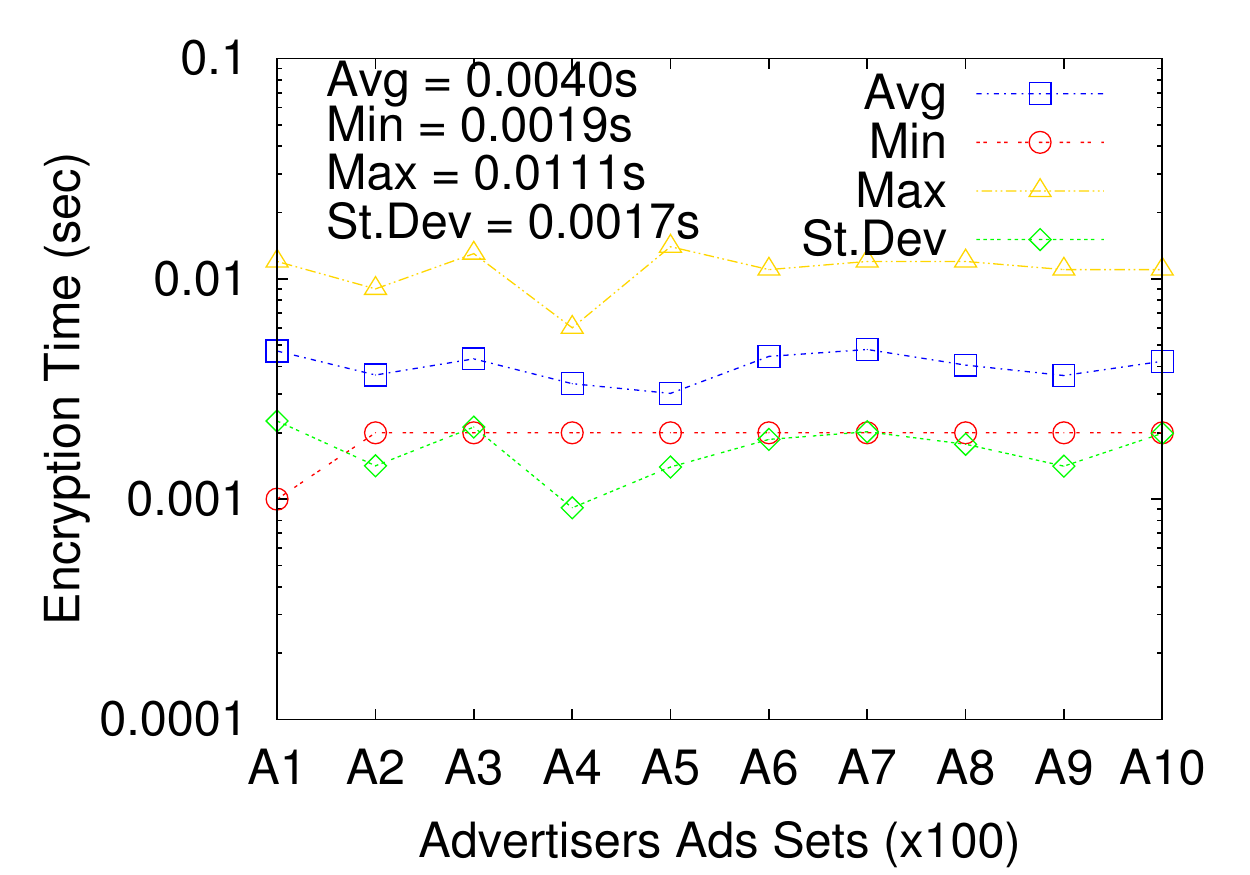}}
      \subfigure[8192]{\includegraphics[scale=0.3]{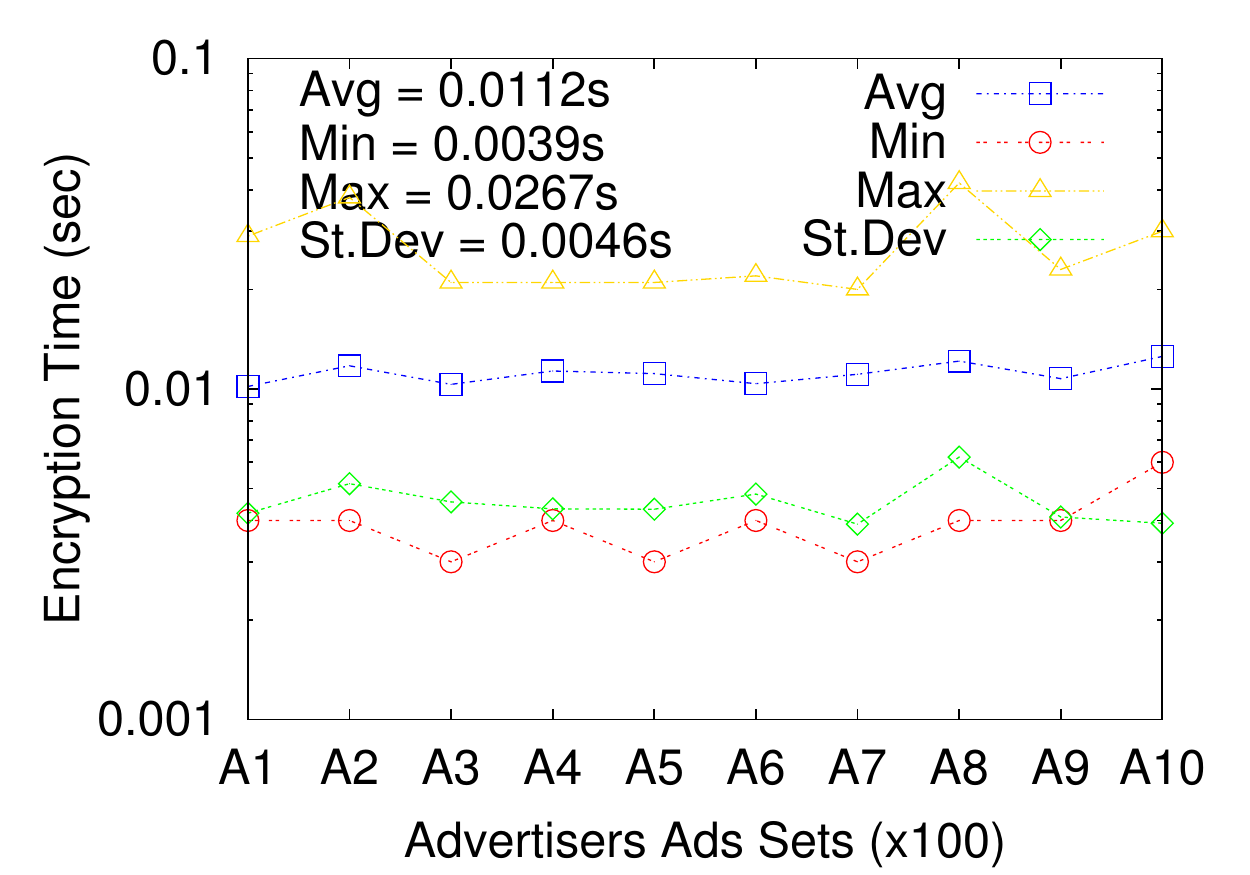}}
     \end{center}
     \begin{center}
      \subfigure[1024]{\includegraphics[scale=0.3]{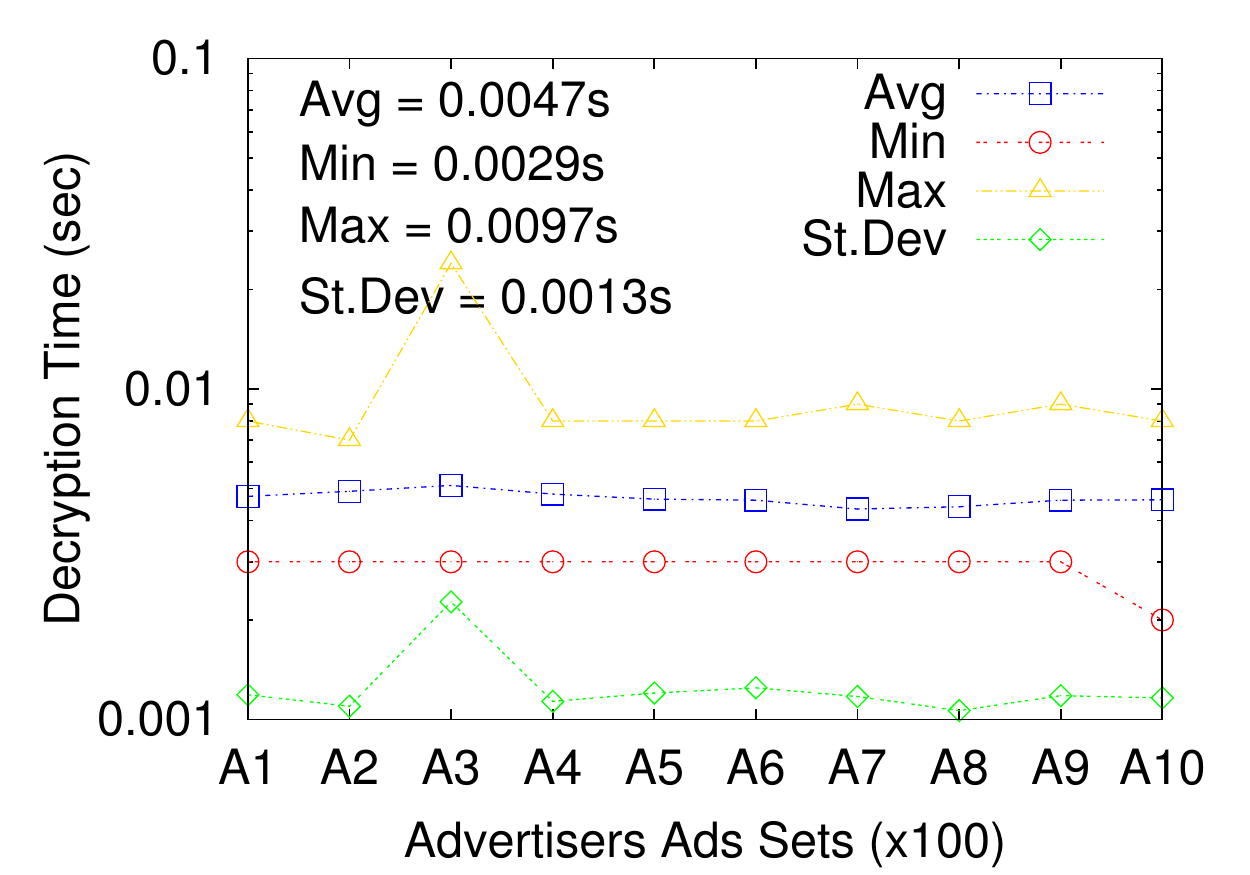}}
      \subfigure[2048]{\includegraphics[scale=0.3]{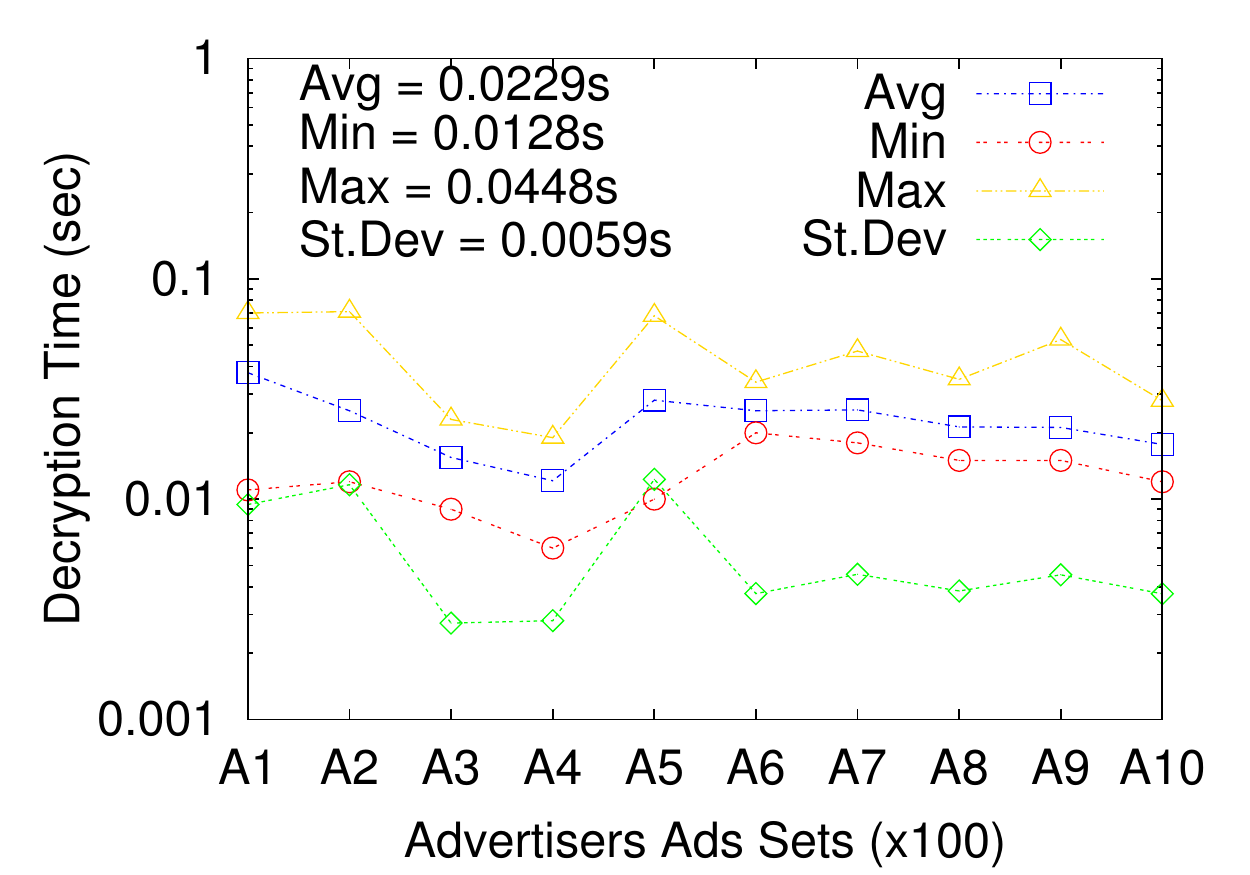}}
      \subfigure[4096]{\includegraphics[scale=0.3]{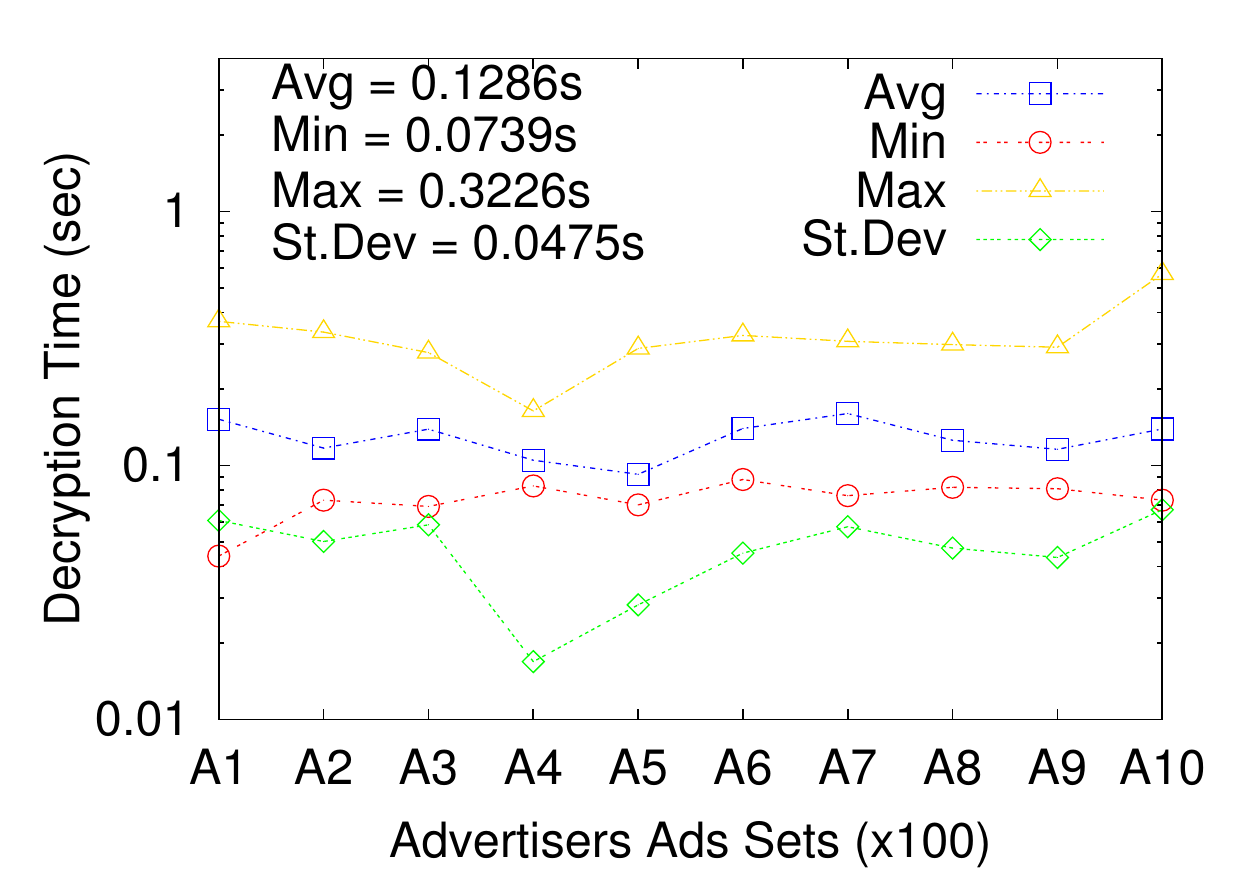}}
      \subfigure[8192]{\includegraphics[scale=0.3]{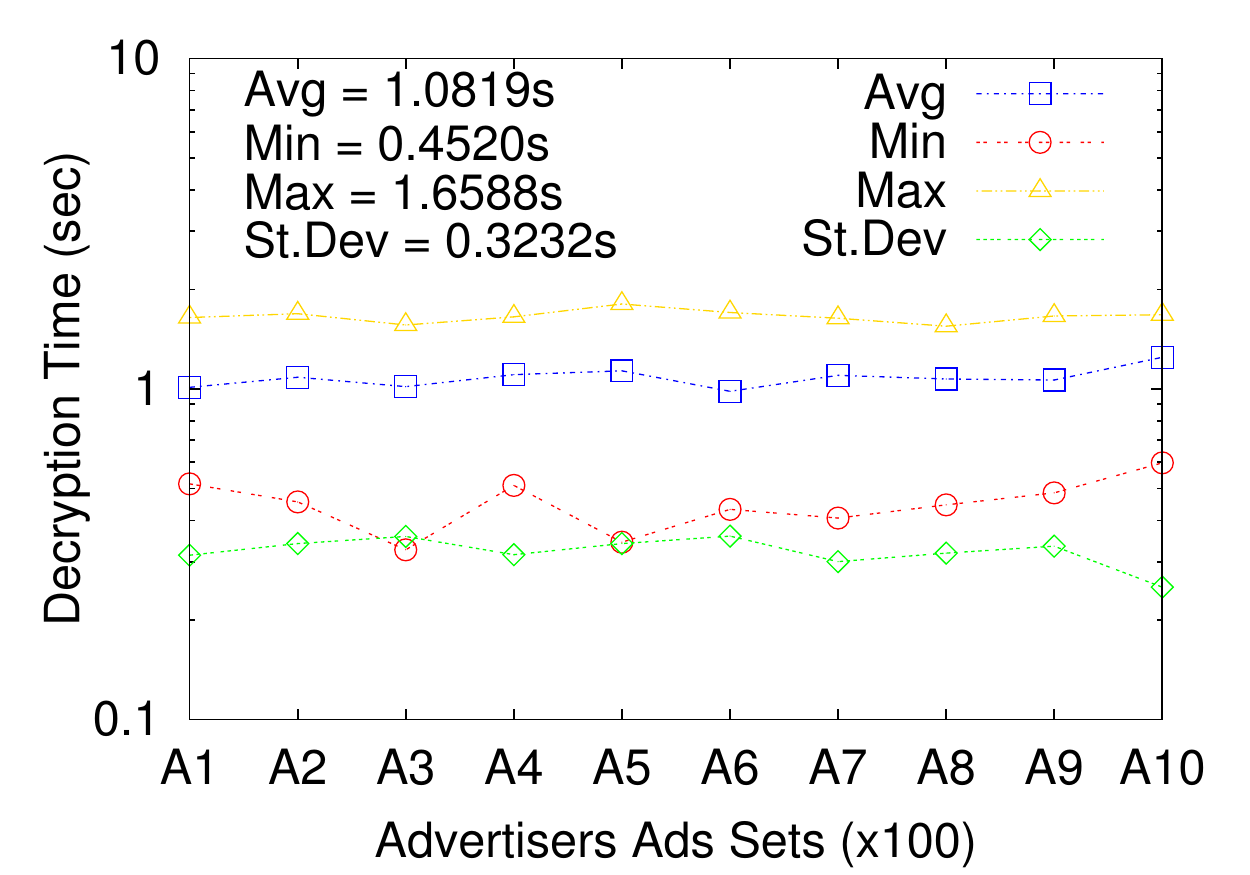}}
     \end{center}
    \caption{\texttt{AES} Encryption time (a) to (d) and Decryption time (e) to (h) using \(1024,2048,4096,8192\) bits of \texttt{RSA} $PK+$ key, evaluated for \({A_1},{A_2}, \ldots ,{A_{10}}\) advertiser's ads groups.}
     \label{ads-enc-dec-time}
\end{figure*}

\begin{table*}[h]
%\begin{center} \scalebox{0.5}{
\begin{center} \scalebox{0.80}{
{
% Preview source code for paragraph 0
\begin{tabular}{|l|c|c|c|c|c|c|c|c|c|c|c|c|c|}
\hline
\multirow{3}{*}{\textbf{Stats}} & \multicolumn{6}{c|}{\textbf{Encryption Time (sec)}} & \multirow{7}{*}{} & \multicolumn{6}{c|}{\textbf{Decryption Time (sec)}}\tabularnewline
\cline{2-7} \cline{9-14}
 & \multicolumn{4}{c|}{\textbf{Key Size (bits)}} & \multicolumn{2}{c|}{\textbf{Trendline}} &  & \multicolumn{4}{c|}{\textbf{Key Size (bits)}} & \multicolumn{2}{c|}{\textbf{Trendline}}\tabularnewline
\cline{2-7} \cline{9-14}
 & \textbf{1024} & \textbf{2048} & \textbf{4096} & \textbf{8192} & \textbf{$R^2$-Exp} & \textbf{$R^2$-Pow} &  & \textbf{1024} & \textbf{2048} & \textbf{4096} & \textbf{8192} & \textbf{$R^2$-Exp} & \textbf{$R^2$-Pow}\tabularnewline
\cline{1-7} \cline{9-14}
\textbf{Avg} & 0.0022 & 0.0031 & 0.0040 & 0.0112 & 0.8921 & 0.7663 &  & 0.0047 & 0.0229 & 0.1286 & 1.0819 & 0.9953 & 0.9320\tabularnewline
\cline{1-7} \cline{9-14}
\textbf{Min} & 0.0010 & 0.0010 & 0.0010 & 0.0030 & 0.6000 & 0.4307 &  & 0.0020 & 0.0060 & 0.0440 & 0.3260 & 0.9839 & 0.8984\tabularnewline
\cline{1-7} \cline{9-14}
\textbf{Max} & 0.0060 & 0.0120 & 0.0140 & 0.0420 & 0.9223 & 0.8496 &  & 0.0240 & 0.0710 & 0.5670 & 1.8070 & 0.9840 & 0.9323\tabularnewline
\cline{1-7} \cline{9-14}
\textbf{St.Dev} & 0.0007 & 0.0014 & 0.0018 & 0.0046 & 0.9627 & 0.9008 &  & 0.0013 & 0.0097 & 0.0534 & 0.3308 & 0.9992 & 0.9690\tabularnewline
\hline
\end{tabular}}}
 \end{center} \caption{The $Avg,Min,Max,St.Dev$ of \texttt{AES} encryption and description processing time evaluated from 1000 advertiser's ads of various sizes with different $PK+$ sizes of \(1024,2048,4096,8192\) bits.\label{ads-enc-dec-time-table}}

\end{table*}

\paragraph*{\textbf{Processing time of Hashing of user profiles}}
Figure \ref{key-user-profiles}(b) shows the processing time of hashing various user profiles with \texttt{SHA1}, \texttt{SHA224}, \texttt{SHA256}, \texttt{SHA384}, \texttt{SHA512} hashing schemes, showing the $Avg,Min,Max,St.Dev$ processing time (msec) of \({P_1},{P_2}, \ldots ,{P_{10}}\) profile groups with different hashing schemes. For instance, the $Min$ and $Max$ hashing time with \texttt{SHA1} are respectively 3msec and 15msec with an $Avg$ and $St.Dev$ of 7.13msec and 1.02ms, while these statistics respectively increase to 5msec, 36msec, 8.19msec and 1.66msec using \texttt{SHA512} hashing scheme. The $Avg$ processing time with each scheme is also shown on top of each schemes. Note that this is a one time operation, which \texttt{APS} calculates before calculating the profile hashing and their relevant ads before \texttt{profiling-ads} to \texttt{CS}.

\paragraph*{\textbf{Enryption/decryption time of Advertiser's ads}}
Figure \ref{ads-enc-dec-time} shows the \texttt{AES} encryption (above row from (a) to (d)) and \texttt{AES} decryption (below row from (e) to (h)) time of various set of ads with different $PK+$ key sizes i.e. \(1024,2048,4096,8192\) bits. The averages of $Avg,Min,Max,St.Dev$ processing time duration of \({A_1},{A_2}, \ldots ,{A_{10}}\) advertiser's ads groups is also shown within each figure for both ad's encryption time (sec) and decryption time (sec) with each evaluation. For instance, the $Min$ and $Max$ ad's encryption time with $1024$ bits key size are respectively 0.0010sec and 0.0045sec with an $Avg$ and $St.Dev$ of 0.0022sec 0.0007sec; on the other hand, with $8192$ bits of key size, these statistics increase to 0.039sec and 0.0267sec respectively for $Min$ and $Max$ encryption time with $Avg$ and $St.Dev$ of 0.0112msec and 0.0046sec. Note that \texttt{APS} needs to encrypt ads only once i.e., at the time of system initialization, as described in Algorithm \ref{algo-profiling}, before uploading to \texttt{CS}. Similarly, the decryption time (below row of Figure \ref{ads-enc-dec-time}) with $Min$ and $Max$ respectively evaluates to 0.0029sec and 0.0097sec with $Avg$ and $St.Dev$ of 0.0047sec and 0.0013sec with $1024$ bits of $PK+$ key size; on the other hand, with $8192$ bits of key size, these statistics evaluate to 0.4520sec, 1.6588sec, 1.0819sec, and 0.3232sec respectively for $Min,Max,Avg$ and $St.Dev$.

Various statistics of $Avg,Min,Max,St.Dev$ of \texttt{AES} encryption and description processing time evaluated from 1000 advertiser's ads with various $PK+$ sizes of \(1024,2048,4096,8192\) bits is presented in Table \ref{ads-enc-dec-time-table}. Furthermore, the $r-squred$ values calculated using \textit{exponential trendline} $R^2-Exp$ and power $R^2-Pow$ are also presented that shows the relationship of increasing key size to the encryption/decryption processing time of advertiser's ads. We note that there is a strong correlation between the $PK+$ key size and the ads' encryption or decryption processing time. For instance, $Avg$ \texttt{AES} encryption processing time, the $R^2-Exp$ and power $R^2-Pow$ evaluated respectively to 0.8921 and 0.7663 (which also evident with high values of $St.Dev$ for both $R^2-Exp$ and $R^2-Pow$) showing that the encryption time exponentially increases as the key size is increased from 1024 to 8192.

\begin{figure*}[ht]
     \begin{center}
      \subfigure[Processing delay (ms)]{\includegraphics[scale=0.5]{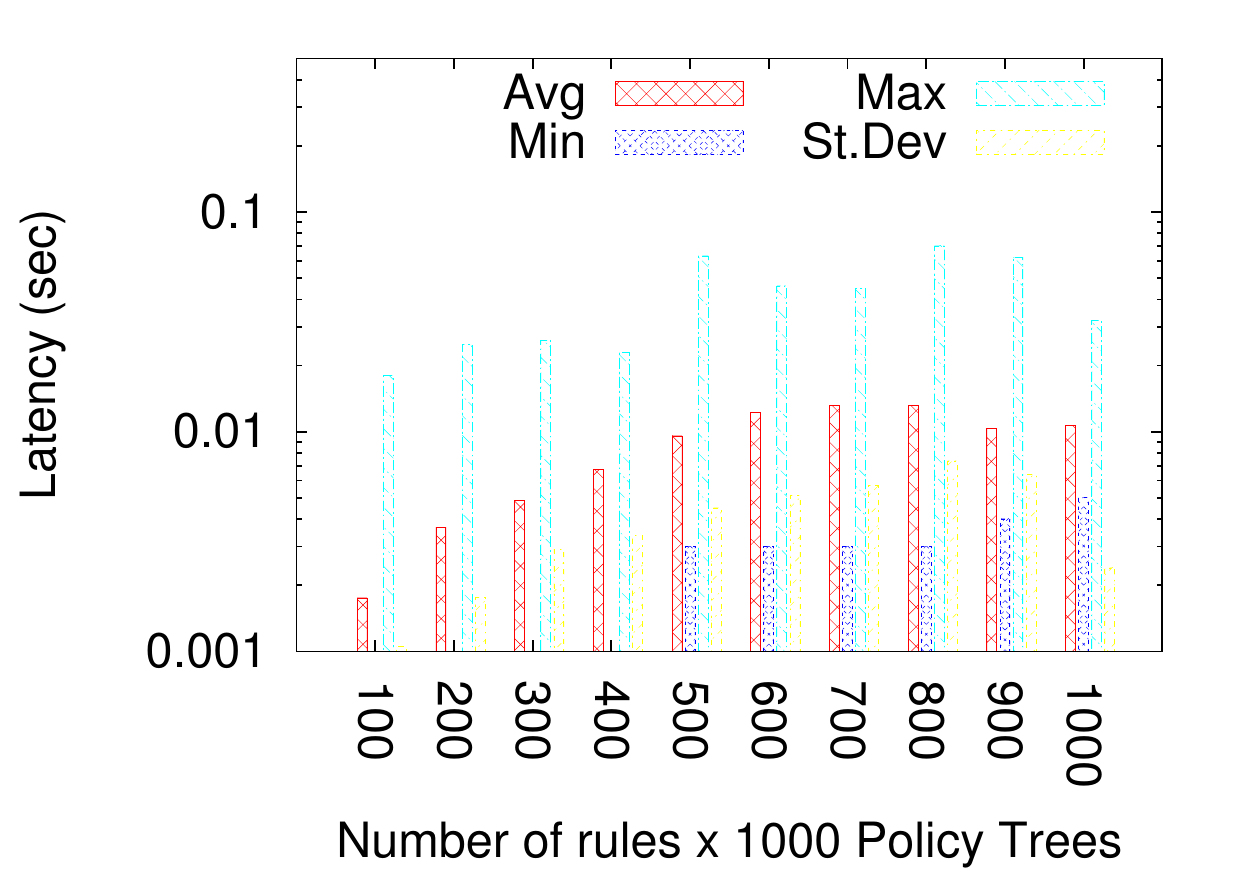}}
      \subfigure[Root node selection]{\includegraphics[scale=0.5]{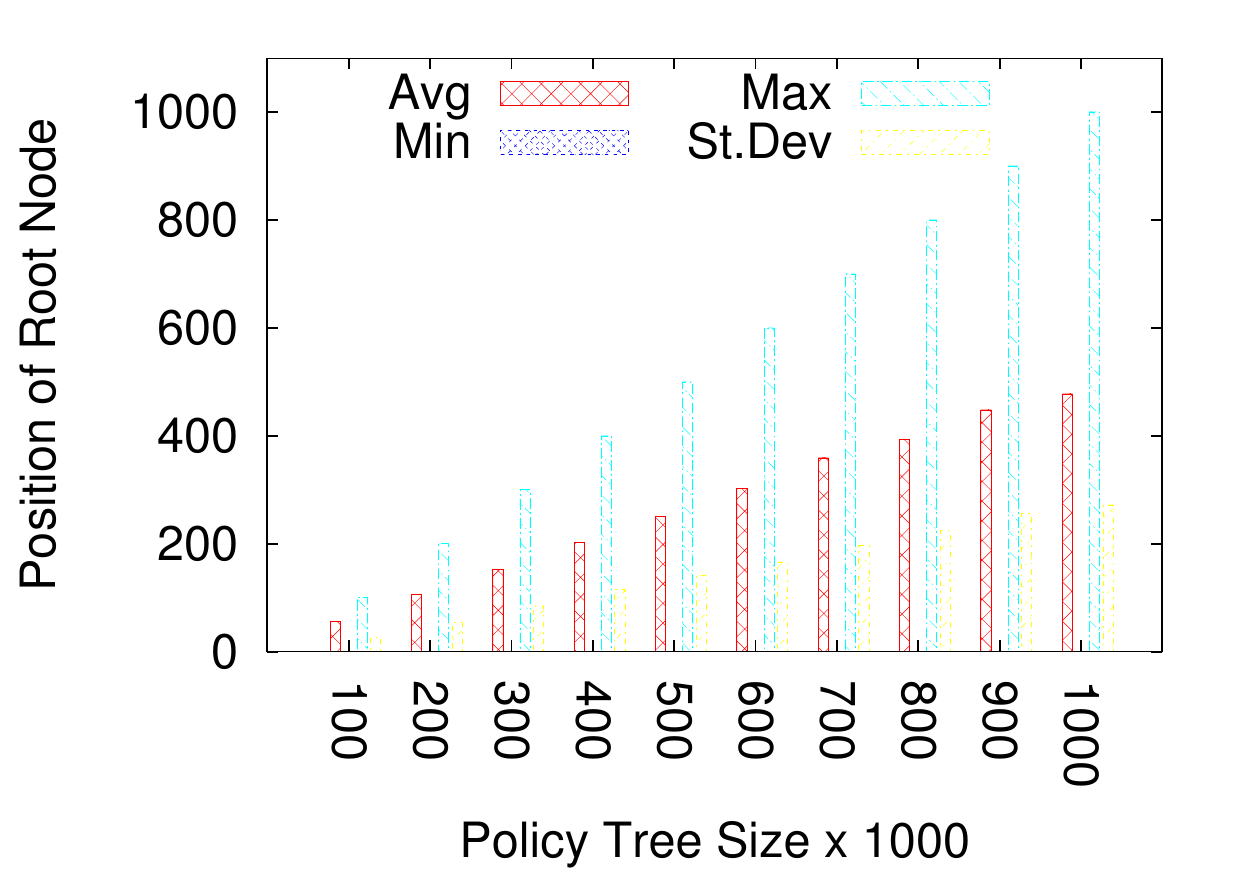}}
     \end{center}
    \caption{$Avg,Min,Max,St.Dev$ Processing delay (ms) per \textit{policy tree} vs number of rules traversed (x1000) for each \textit{policy matching} requested by \textit{Miner} (a) and root node placement for each set of \textit{policy trees} (b).}
     \label{policy-tree}
\end{figure*}

% Preview source code for paragraph 0

\begin{table*}[h]
%\begin{center} \scalebox{0.5}{
\begin{center} \scalebox{0.80}{
{
% Preview source code for paragraph 0
\textbf{}%
\begin{tabular}{|l|c|c|c|c|c|c|c|c|c|}
\hline
\multirow{2}{*}{\textbf{Tree Size}} & \multicolumn{4}{c|}{\textbf{Policy Tree Processing Delay (sec)}} & \multirow{14}{*}{} & \multicolumn{4}{c|}{\textbf{Policy Tree - Rood Node Selection}}\tabularnewline
\cline{2-5} \cline{7-10}
 & \textbf{Avg} & \textbf{Min} & \textbf{Max} & \textbf{St.Dev} &  & \textbf{Avg} & \textbf{Min} & \textbf{Max} & \textbf{St.Dev}\tabularnewline
\cline{1-5} \cline{7-10}
\textbf{100} & 0.0017 & 0.001 & 0.018 & 0.001 &  & 56.08 & \multirow{10}{*}{1} & 100 & 26.95\tabularnewline
\cline{1-5} \cline{7-7} \cline{9-10}
\textbf{200} & 0.0037 & 0.001 & 0.025 & 0.002 &  & 106.63 &  & 200 & 53.63\tabularnewline
\cline{1-5} \cline{7-7} \cline{9-10}
\textbf{300} & 0.0049 & 0.001 & 0.026 & 0.003 &  & 152.98 &  & 300 & 84.94\tabularnewline
\cline{1-5} \cline{7-7} \cline{9-10}
\textbf{400} & 0.0067 & 0.001 & 0.023 & 0.003 &  & 202.37 &  & 400 & 115.29\tabularnewline
\cline{1-5} \cline{7-7} \cline{9-10}
\textbf{500} & 0.0096 & 0.003 & 0.063 & 0.004 &  & 250.99 &  & 500 & 141.94\tabularnewline
\cline{1-5} \cline{7-7} \cline{9-10}
\textbf{600} & 0.0122 & 0.003 & 0.046 & 0.005 &  & 301.90 &  & 600 & 166.34\tabularnewline
\cline{1-5} \cline{7-7} \cline{9-10}
\textbf{700} & 0.0131 & 0.003 & 0.045 & 0.006 &  & 359.22 &  & 700 & 196.66\tabularnewline
\cline{1-5} \cline{7-7} \cline{9-10}
\textbf{800} & 0.0131 & 0.003 & 0.071 & 0.007 &  & 393.72 &  & 800 & 224.08\tabularnewline
\cline{1-5} \cline{7-7} \cline{9-10}
\textbf{900} & 0.0103 & 0.004 & 0.062 & 0.006 &  & 447.85 &  & 900 & 255.95\tabularnewline
\cline{1-5} \cline{7-7} \cline{9-10}
\textbf{1000} & 0.0107 & 0.005 & 0.032 & 0.002 &  & 477.54 &  & 1000 & 271.96\tabularnewline
\cline{1-5} \cline{7-10}
\textbf{$R^2-Pow$} & \textbf{0.9166} & \textbf{0.7491} & \textbf{0.5778} & \textbf{0.6673} &  & \textbf{0.9994} & \textbf{0.0436} & \textbf{0.9996} & \textbf{0.9991}\tabularnewline
\cline{1-5} \cline{7-10}
\textbf{$R^2-Exp$} & \textbf{0.7196} & \textbf{0.8497} & \textbf{0.4804} & \textbf{0.4553} &  & \textbf{0.9084} & \textbf{0.0303} & \textbf{0.8985} & \textbf{0.8940}\tabularnewline
\hline
\end{tabular}
}}
 \end{center} \caption{$Avg,Min,Max,St.Dev$ \textit{Policy tree} processing delay (sec) (left) and the \textit{random} root node placement position (right) for \textit{policy trees} of $100, 200, \ldots , 1000$; the \textit{exponential trendline} $R^2-Exp$ and power $R^2-Pow$ are also presented. Each statistic is a result of 1000 experimental request.\label{tree-node-processing-delay}}

\end{table*}

\begin{figure}[h]
\begin{center}
\includegraphics[scale=0.7]{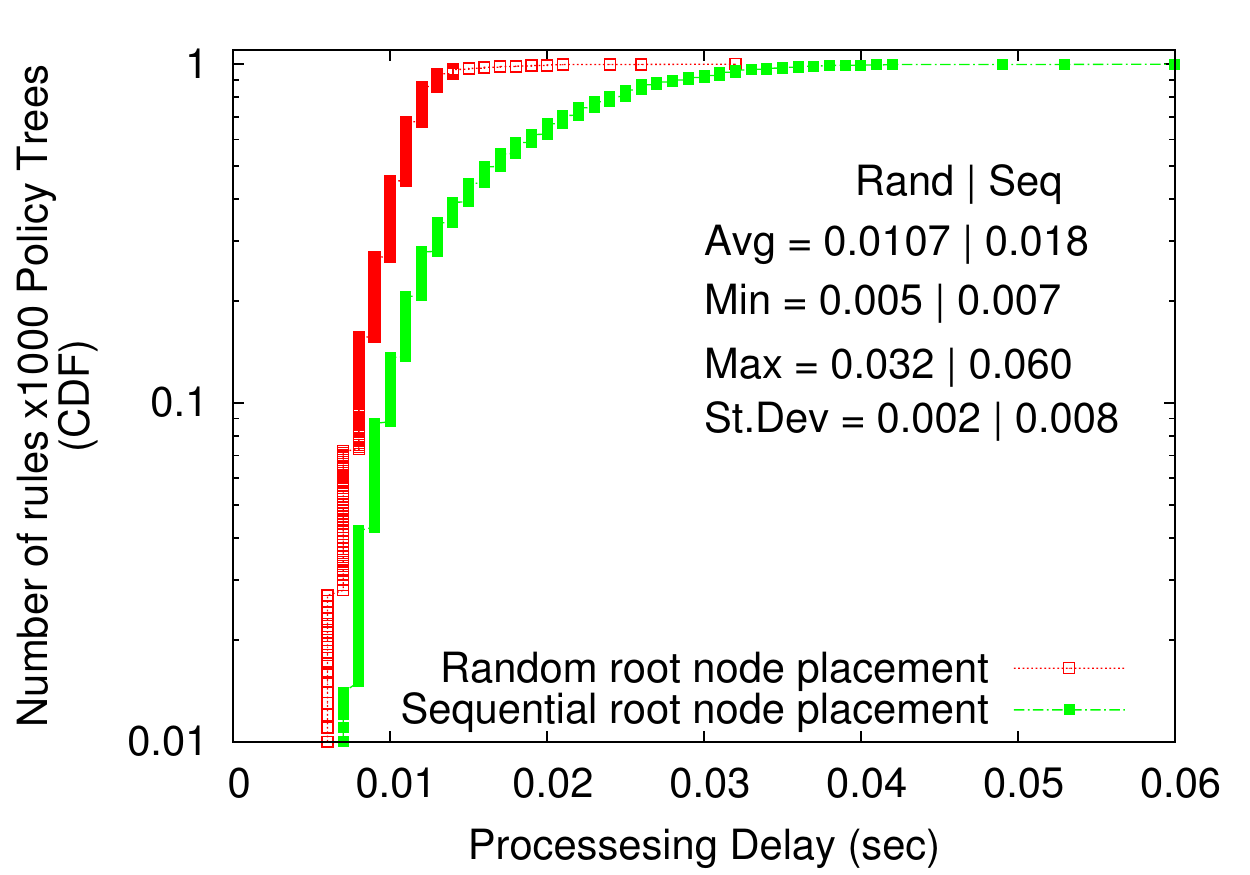}
\caption{$Avg,Min,Max,St.Dev$ Processing delay (ms) per \textit{policy tree} vs number of rules traversed for rules matching requested by \textit{Miner} for \textit{Random} and \textit{Sequential} root nodes placement.}
\label{tree-rand-geom-latency}
\end{center}
\end{figure}

\paragraph*{\textbf{Processing delay for traversing policies}}
To further analyze the scalability of the our proposed framework, we study the processing delay (note that this contributes to the total delay of various transactions requested by \textit{Miner} or any other entity that requests \texttt{CS} for various services) of $Avg,Min,Max,St.Dev$ when varying the number of traversed rules among $100, 200, \ldots , 1000$ \textit{access policy trees}, for both \textit{sequential} and \textit{random} distribution of \textit{policy tree} i.e. \textit{sequential/random} root node placement. Figure \ref{policy-tree} (b) shows the \textit{random} placement of root node for \textit{policy trees}, storing various nodes and matching rules of $100, 200, \ldots , 1000$ access \textit{policy trees}. We note that the root node placement, on average, perfectly follows a power distribution with $R^2-Pow$ value of 0.99, as the tree is populated with 100 to 1000 policies and matching rules. Similarly, the $R^2-Pow$ for $Max,St.Dev$ is also 0.99, suggesting formation of various \textit{access policy} with placement of root nodes at every possible positions i.e., creating every possible shape of tree by placing nodes on both sides of the access \textit{policy trees}. %Note that the $Min$ and $Max$ values for root node placement in every possible access trees is respectively at positions 1 and 1000, forming a tree where the entire nodes (matching rules and corresponding actions) are respectively placed on right and left side of the root node.

Figure \ref{policy-tree} (a) shows processing delay of $Avg,Min,Max,St.Dev$ \textit{access policy} traversal delay (msec) with various access trees; note that every experiment has been repeated 1000 times. We observe that the average processing delay increases with the number of rules with a \textit{power trendline} of 0.92. For instance, for a 100 and 1000 node's \textit{policy tree}, the $Avg$ processing delay is respectively 0.0017sec and 0.0106sec, similarly, the $Min$ processing delay varies respectively from 0.001sec to 0.005sec. Detailed statistics of root node placement and traversal delay is also given in Table \ref{tree-node-processing-delay}, evaluated for $Avg,Min,Max,St.Dev$ processing delays.

Following, we evaluate the processing delay using \textit{sequential} (distribution) placement of root node (i.e. to evaluate traversal delay by placing root node at every single position e.g. $1,2,3,\ldots,1000$ for a tree of 1000 matching rules and their actions) and compare it with the \textit{random} distribution. Figure \ref{tree-rand-geom-latency} presents a CDF of processing delay with both \textit{random} and \textit{sequential} placement of rood node; various processing delays of $Avg,Min,Max,St.Dev$ is also given at the right side of this figure. We note that 96\% of the processing delay ranges from 0.005sec to 0.014sec and 0.007sec to 0.032sec respectively for \textit{random} and \textit{sequential} distribution of root node placement. Overall, the \textit{sequential} distribution resulted in higher processing delays as compared to \textit{random} distribution. Note that these processing delay at \texttt{CS} can be considered acceptable i.e., close to real time performance \cite{nath2015madscope, ullah2017enabling}, both during the app/website \textit{launch time} or \textit{delayed} ad request.

%% file: related-work.tex
Privacy threats resulting from collecting individual's online data i.e. direct and indirect (inferred) leakage, have been investigated extensively in literature, e.g. \cite{mamais2019privacy, englehardt2016online, frik2020impact, meng2016price}, including third party ad tracking and visiting \cite{shuba2020nomoats, iqbal2020adgraph, merzdovnik2017block, das2018web}. These studies have shown the prevalence of tracking on both web and mobile environments and demonstrate the possibility of inferring user's PII, such as age, gender, relationship status, email address, etc. from their online data. The ad and analytics libraries, integrated into mobile apps, systematically collect personal information and users' \textit{in-app} activities and are more likely to leak this information to ad systems and third party tracking. The authors in \cite{liu2019privacy} analyze the privacy policy for collecting \textit{in-app} data by apps and study various information collected by the analytics libraries integrated in mobile apps.

\paragraph*{\textbf{Privacy and mobile advertising}}
There are number of studies that focus on privacy leakage caused by ad libraries within the mobile apps, e.g. the authors in \cite{grace2012unsafe} show the potential privacy and security risk by analyzing 100,000 android apps and find that majority of the embedded libraries collect private information. Another study \cite{book2013case} finds that majority of the used APIs within mobile apps do not implement private APIs and send private information to ad servers. The authors in \cite{book2015empirical} manually create user profiles and find that ads are vastly targeted based on fabricated profiles, in addition to, apps used and location. In our previous study \cite{ullah2014characterising} and in \cite{nath2015madscope}, the authors study various information sent to ad networks and the level of ads targeting based on communicated information, similarly, \cite{demetriou2016free} investigates installed apps for leaking targeted user data. The authors in \cite{taylor2017intra} studies the privacy risks collected by ad libraries across various apps installed on a single device called intra-library collusion; the longitudinal analysis shows that leakage of information has increased in the following two years. Furthermore, \cite{meng2016price} studies the level of information collected by the ad networks with the help of installed and used apps, in addition, to reverse engineer the targeted ads and investigate the collected information.

\paragraph*{\textbf{Privacy and mobile Analytics services}}
In our previous work \cite{tchen2014}, we study the user's sensitive information leakage through the vulnerabilities in mobile analytics services by analyzing actual network traffic generated by mobile apps and reconstruct the exact user profiles of a number of participating volunteers. We mainly carry out this study on android platform for Google Mobile Analytics and Flurry Analytics. We further study how the ads served to users can be influenced by modifying the user profiles generated by these analytics services. Another study \cite{seneviratne2015measurement} studies third-party analytics and ad libraries, and find that 60\% of the paid apps contain at least one tracker that collects personal information. A similar study \cite{han2012study} examines how users are tracked by running android apps; they find that 57\% of the apps and every participant in their study is tracked several times for sensitive information. The authors in \cite{vallina2016tracking} collects network traffic using ` ICSI Haystack' app and distinguishes between ad and analytics libraries and for collecting various information. Furthermore, the authors in \cite{binns2018measuring, binns2018third} study the distribution of third-party trackers across web and android apps and their impact on user privacy.

\paragraph*{\textbf{Analysis of third-party tracking scripts}}
Third-party web tracking, the practice using which websites identify and collect information about users, has been widely studied in literature, e.g. \cite{roesner2012detecting} develops a client side and identify 500 unique trackers and finds that majority of commercial websites are tracked by multiple parties. The authors in \cite{lerner2016internet} develop a tool TrackingExcavator' for longitudinal measurements of third-party web tracking 1996-2016 and finds that tracking mechanism is constantly becoming complex and widely used. An open-source\footnote{\url{https://github.com/mozilla/OpenWPM}} privacy measurement tool `OpenWPM' automatically detects and characterizes emerging online tracking behaviors. The `TrackingFree' \cite{pan2015not} provides an anti-tracking browser solution that automatically breaks the tracking chain for third-party tracking web tracking by disabling unique identifies. Similarly, \cite{merzdovnik2017block} analyze the protection level offered across a wide range of web and mobile apps and evaluates their effectiveness for blocking third-party tracking. They find that more than 60\% of third-party tracking services communicate in plain text.

\paragraph*{\textbf{Privacy-preserving mobile advertising}}
There are a number of privacy-preserving personalization and architectures both for in-browser and \textit{in-app} that propose private profiling and advertising systems, such as, Adnostic \cite{toubiana2010adnostic}, Privad \cite{guha2011privad}, RePriv \cite{fredrikson2011repriv}, MobiAd \cite{haddadi2010mobiad}, Splitx \cite{Chen:2013:Splitx}. Adnostic, Privad and RePriv are browser-based extensions that focus on targeting users based on their browsing activities and locally carry out profiling (i.e. in the user's browser). MobiAd suggest local user profiling and receives ads in Delay Tolerant Networks (DTN) fashion via broadcasting stations and WiFi hotspots. Similarly, SplixX uses differentially private queries over distributed user data. Furthermore, another noisy technique `Bloom cookies' \cite{mor2015bloom} presents a framework for privacy preserving personalization of web searches of locally derived profile. In our previous works, we propose `ProfileGuard' \cite{ullah2014profileguard, ullah2020protecting} that is an app-based profile obfuscation mechanism and a privacy-preserving mobile advertising system based on various implementation of Private Information Retrieval (PIR) \cite{ullah2017enabling}.

The authors in \cite{gu2018secure} present a targeted mobile coupon delivery scheme based on Blockchain, however this framework does not include all components of the advertising system including user profile construction, detailed structure of various transactions and operations, or other entities from our work, such as the \textit{Miner} and the \textit{billing} process. In addition, it does not specify the technical details of \textit{access policy} to control access to resources, e.g., ads. Another work \cite{zyskind2015decentralizing} implements a protocol for personal data management using Blockchain that ensures user's ownership and control over data. We present a decentralized Blockchain-based advertising system that ensures private user profiling, in addition, the transfer of information between user and \texttt{CS} and between \texttt{CS} and \texttt{APS} uses an encryption algorithm to ensure data security. Furthermore, the control of information and user's ownership and control over its profiling data is ensured using \textit{access policy}. We also propose private \textit{billing} mechanism for ad presentations and clicks based on Cryptocurrency. For scalability of the system, we introduce the cluster head that can process \textit{ad-block} requests for \textit{Miner} for storing user profiles and fetching ads from \texttt{CS}. 

%% file: conclusion.tex
The \textit{in-app} mobile advertising is growing in popularity, while the advertising companies use excellent user-tracking tools, which have raised concerns among privacy advocates. This paper presents the design of a Blockchain-based framework for advertising platform in addition to implementation of its various critical components that allows the cloud system to privately compute access policies both for mobile users and ad systems for accessing various resources, private user profiling, private \textit{billing}, privately requesting ads from the advertising system based on user interests, and devising various transactions for enabling advertising operations in Blockchain. We develop POC implementation of our proposed framework and perform a thorough system evaluation and stress test its various components in order to evaluate different constraints of processing delays. Our evaluations show an acceptable amount of processing delay for performing various operations of accessing relevant ads as that of the currently implemented ad system in an actual environment.